\documentclass
[superscriptaddress,secnumarabic,amssymb,amsmath,nobibnotes,aps,prd,showkeys,showpacs,nofootinbib,onecolumn]{revtex4}%
\usepackage{graphicx}
\usepackage{epsf}
\usepackage{bm}
\usepackage{amsmath}
\usepackage{amsfonts}
\usepackage{amssymb}%
\providecommand{\U}[1]{\protect\rule{.1in}{.1in}}
\newcommand{\be}{\begin{equation}}
\newcommand{\ee}{\end{equation}}

\newcommand{\mincir}{\raise
-3.truept\hbox{\rlap{\hbox{$\sim$}}\raise4.truept\hbox{$<$}\ }}
\newcommand{\magcir}{\raise
-3.truept\hbox{\rlap{\hbox{$\sim$}}\raise4.truept\hbox{$>$}\ }}

\ifx\pdfoutput\relax\let\pdfoutput=\undefined\fi
\newcount\msipdfoutput
\ifx\pdfoutput\undefined\else
\ifcase\pdfoutput\else
\ifx\paperwidth\undefined\else
\ifdim\paperheight=0pt\relax\else\pdfpageheight\paperheight\fi
\ifdim\paperwidth=0pt\relax\else\pdfpagewidth\paperwidth\fi
\fi\fi\fi
\begin{document}
\title{Cyclic Szekeres Universes}
\author{John D. Barrow}
\email{J.D.Barrow@damtp.cam.ac.uk}
\affiliation{DAMTP, Centre for Mathematical Sciences, University of Cambridge, Wilberforce
Rd., Cambridge CB3 0WA, UK}
\author{Andronikos Paliathanasis}
\email{anpaliat@phys.uoa.gr}
\affiliation{Instituto de Ciencias F\'{\i}sicas y Matem\'{a}ticas, Universidad Austral de
Chile, Valdivia, Chile}
\affiliation{Institute of Systems Science, Durban University of Technology, PO Box 1334,
Durban 4000, RSA}

\begin{abstract}
We consider the Szekeres universe with an inhomogeneous dust fluid and a
homogeneous and isotropic ghost matter source with equation of state
$p_{g}=\left(  \gamma-1\right)  \rho_{g},$ where $\gamma$ is a constant. The
field equations determine two families of spacetimes which describe
homogeneous Kantowski-Sachs universes and inhomogeneous Friedmann universe.
The ghost field Einstein permits static and cyclic solutions to exist. The
stability of the Einstein static and cyclic solutions are studied with a
critical point analysis.

\end{abstract}
\maketitle
\date{\today}

\section{Introduction}

One proposal to solve the flatness and the horizon problems of our universe,
that differs from the inflationary scenario \cite{guth}, is the cyclic
cosmological model \cite{cyc1}. In the cyclic model, the universe undergoes an
endless series of cycles of expansion and contraction, and the cosmic energy
density and cosmic temperature remain finitely defined at any transition
between expanding and contracting phases of the universe.

In the theory of general relativity cyclic universes can be constructed in the
presence of a ghost field \cite{gyb}. Ghost fields are exotic matter sources
with negative energy density and also can have a parameter for the equation of
state $w_{f}$ $=p/\rho$ for pressure $p$ and density $\rho,$ such that
$w_{f}<-1$. There are various applications of ghost fields in classical and
quantum cosmology \cite{s1,s2,s3,s4} and it is interesting to note that the
stability of Einstein static universes changes in the presence of a ghost
field. More specifically, in \cite{jts} it was found that there exact
solutions which describe an oscillation around an Einstein static solution for
a closed Friedmann--Lema\^{\i}tre--Robertson--Walker universe (FLRW) when a
radiation-ghost field ($w_{f}=1/3,\rho<0$) exists. More recently, the
behaviour of cyclic mixmaster universes was studied in \cite{jkm,jkm2} in the
presence of ghost fields.

In this work, we study the existence of ghost fields in inhomogeneous dust
universes \cite{sz00} by assuming a \textquotedblleft silent
universe\textquotedblright\ \cite{silent1} with dust and a radiation-like
ghost matter source. More specifically, we focus on the existence and
stability of Szekeres-like cyclic universes. Szekeres universes
\cite{szek0,szafron} describe exact inhomogeneous solutions in general
relativity which does not admit any isometry {\cite{musta}. These exact
solutions are categorized in two large families of spacetimes, the
inhomogeneous Kantowski-Sachs solutions and the inhomogeneous FLRW solutions.
Various applications of the Szekeres universes can be found in
\cite{sz2,sz2a,sz3,sz4,sz5,sz6,sz7}. A detailed analysis of the conservation
laws and the dynamics of the Szekeres system was performed recently in
\cite{f1,f2}. The results of \cite{f1} were applied in \cite{f3} to
quantize\ the Szekeres system for the first time. }

{A generalization of the Szekeres solutions in the presence of a cosmological
constant was presented in \cite{barcc}; while the inclusion of a fluid source
with heat flow in Szekeres universes was made in \cite{hf1}. }Re{cently, the
case of the Szekeres inhomogeneous dust model with a homogeneous scalar field
was studied in \cite{jdbsc}. In \cite{jdbsc} it was found that there exists
only one family of solutions which describe inhomogeneous universes and they
generalise the FLRW family. By contrast, the Szekeres family of solutions of
Kantowski-Sachs type describe spatially homogeneous universes when the ghost
field is added to the dust. }

In the following, we consider the Szekeres system with a dust fluid and a
homogeneous ghost matter source with constant parameter for the equation of
state. We solve the gravitational field equations analytically and we find
that the two families of solutions are those of homogeneous Kantowski-Sachs
and inhomogeneous FLRW spacetimes. These results are similar to that of the
Szekeres model with a homogeneous scalar field {\cite{jdbsc}}. For the
inhomogeneous FLRW-like solution we are able to write the solution in a closed
form. More specifically, we find again that for a closed FLRW-like universe
there exists a periodic solution around a static universe. Furthermore, from
the stability analysis we find that all the solutions in which the expansion
rate $\theta$ changes sign are unstable. We perform that analysis by studying
the field equations in dimensionless variables different from those of the
$H-$normalization \cite{silent1}.

The plan of the paper is as follows. In Section \ref{section2s} we define our
cosmological model which is that of the Szekeres metric with a homogeneous and
isotropic ghost field with constant parameter for the equation of state. The
requirement of homogeneity for the ghost field provides a first constraint on
the unknown functions in the line element for the geometry of the universe. In
Section \ref{section3} we present the two families of spacetimes which
describe the solutions of the field equations. The stability of the cyclic
solutions is presented in Section \ref{cyc1}. Finally, in Section
\ref{section5} we discuss our results and we draw our conclusions.

\section{Szekeres universes with dust and an isotropic ghost field}

\label{section2s}

In the context of general relativity we consider the action integral of the
field equations to be%
\begin{equation}
S=\int d^{4}x\sqrt{-g}R+\int d^{4}x\sqrt{-g}L_{m}+\int d^{4}x\sqrt{-g}%
L_{g},\label{ee.01}%
\end{equation}
where $L_{m}$ is the Lagrangian density of a pressureless fluid term and
$L_{G}$ describes an isotropic and homogeneous ghost ideal gas.

The Einstein field equations are%
\begin{equation}
G_{\mu\nu}=T_{\mu\nu}^{\left(  m\right)  }+T_{\mu\nu}^{\left(  g\right)  }
\label{ee.02}%
\end{equation}
in which
\begin{equation}
T^{\mu\nu\left(  m\right)  }=-\frac{1}{2\sqrt{-g}}\frac{\partial\left(
\sqrt{-g}L_{m}\right)  }{\partial g_{\mu\nu}}~~\text{and ~}T^{\mu\nu\left(
g\right)  }=-\frac{1}{2\sqrt{-g}}\frac{\partial\left(  \sqrt{-g}L_{g}\right)
}{\partial g_{\mu\nu}}. \label{ee.03}%
\end{equation}
where the Bianchi identity gives $\left(  T^{\mu\nu\left(  m\right)  }%
+T^{\mu\nu\left(  g\right)  }\right)  _{;\nu}=0$. \ Furthermore, by assuming
that the two matter sources (dust and ghost field) are minimally coupled, we
end up with two separate conservation equations:%
\begin{equation}
\left(  T^{\mu\nu\left(  m\right)  }\right)  _{;\nu}=0~,~\left(  T^{\mu
\nu\left(  g\right)  }\right)  _{;\nu}=0. \label{ee.04}%
\end{equation}

For the background metric, we consider the following line element introduced
by Szekeres \cite{szek0}:%
\begin{equation}
ds^{2}=-dt^{2}+e^{2A}dr^{2}+e^{2B}\left(  dy^{2}+dz^{2}\right)  ,
\label{ee.05}%
\end{equation}
where functions of $A=A\left(  t,r,y,z\right)  $ and $B=B\left(
t,r,y,z\right)  $ are solutions of the Einstein's field equations (\ref{ee.02}).

In terms of $1+3$ decomposition for the fluid sources we have%
\begin{equation}
T_{\mu\nu}^{\left(  m\right)  }=\rho_{m}\left(  t,r,y,z\right)  u_{\mu}u_{\nu
}~\ ,~T_{\mu\nu}^{\left(  g\right)  }=\rho_{g}\left(  t\right)  u_{\mu}u_{\nu
}+p_{g}\left(  t\right)  h_{\mu\nu}, \label{ee.06}%
\end{equation}
where $u^{\mu}=\delta_{t}^{\mu}$ is the comoving 4-velocity and $h_{\mu\nu
}=g_{\mu\nu}+u_{\mu}u_{\nu}$ is the projective tensor, $\rho_{m}$ is the
inhomogeneous dust density, and for the homogeneous ghost field we set
$p_{g}\left(  t\right)  =\left(  \gamma-1\right)  \rho_{g}\left(  t\right)
.$Both fluids share the same 4-velocity.

By substituting (\ref{ee.06}) into (\ref{ee.04}), we find
\begin{equation}
\frac{\partial\rho_{m}\left(  t,r,y,z\right)  }{\partial t}+\left(
\frac{\partial A\left(  t,r,y,z\right)  }{\partial t}+2\frac{\partial B\left(
t,r,y,z\right)  }{\partial t}\right)  \rho_{m}\left(  t,r,y,z\right)  =0,
\label{ee.07}%
\end{equation}%
\begin{equation}
\frac{\partial\rho_{g}\left(  t\right)  }{\partial t}+\gamma\left(
\frac{\partial A\left(  t,r,y,z\right)  }{\partial t}+2\frac{\partial B\left(
t,r,y,z\right)  }{\partial t}\right)  \rho_{g}\left(  t\right)  =0
\label{ee.08}%
\end{equation}
which implies \cite{jdbsc}:%
\begin{equation}
\exp\left(  A\left(  t,r,y,z\right)  \right)  =a\left(  t\right)  \exp\left(
F\left(  r,y,z\right)  -2B\left(  t,r,y,z\right)  \right)  . \label{ee.09}%
\end{equation}

We proceed now with the presentation of the possible solutions for the
Einstein field equations (\ref{ee.02}).

\section{Analytic cyclic solutions}

\label{section3}

Szekeres spacetimes correspond to two families, the Kantowski-Sachs family
with $\frac{\partial B}{\partial r}=0$ and the FLRW family in which
$\frac{\partial B}{\partial r}\neq0$. While in the Szekeres system the two
spacetimes are inhomogeneous and do not admit any isometry, in \cite{jdbsc} it
was found that, if an isotropic scalar field is added to the dust source then
the Kantowski-Sachs family of solutions must be spatially homogeneous, while
an extra constraint on the functional form of the spacetime appears for the
inhomogeneous FLRW family. In a similar way, the same two families of
solutions are determined for the model considered here.

In particular, for the homogeneous Kantowski-Sachs family, the line element
(\ref{ee.05}) simplifies to%
\begin{equation}
ds^{2}=-dt^{2}+a^{2}\left(  t\right)  dr^{2}+b^{2}\left(  t\right)
\frac{\left(  dy^{2}+dz^{2}\right)  }{\left(  c_{1}\left(  \left(
y-y_{0}\right)  ^{2}+\left(  z-z_{0}\right)  ^{2}\right)  +c_{2}\right)  ^{2}%
},\label{sf.11a}%
\end{equation}
with $c_{1},c_{2}$ constants, while the gravitational field equations reduce
to those of the Kantowski-Sachs spacetime with two homogeneous perfect fluids
\cite{ks011} whose solution gives the evolution of the scale
factors\footnote{Note that in the case of homogeneous perfect \ fluids, for
the line element \ref{sf.11a} the conservation equations (\ref{ee.07}%
)-(\ref{ee.08}) give $\rho_{m}(t)=\rho_{m0}a^{-1}b^{-2}$ and $\rho_{g}%
=\rho_{g0}a^{-\gamma}b^{-2\gamma}$, in which $\rho_{mo}$ and $\rho_{go}$ are
constants of integration.} $a(t)$ and $b(t)$. Moreover, the spatial curvature
$K$ of the 2-dimensional line element $ds_{\left(  2\right)  }^{2}=\left(
c_{1}\left(  \left(  y-y_{0}\right)  ^{2}+\left(  z-z_{0}\right)  ^{2}\right)
+c_{2}\right)  ^{-2}\left(  dy^{2}+dz^{2}\right)  $ is calculated to
be\footnote{When $K=0$, the line element (\ref{sf.11a}) describes the
homogeneous Bianchi I spacetime, while, when $K>0$\thinspace, the line element
(\ref{sf.11a}) is that of the Bianchi III spacetime.} $K=8c_{1}c_{2}$.

The nonlinearity of the field equations prevents us from finding closed-form
solutions. However, for $K=0$ (or in the limit in which$~\frac{K}{b^{2}%
}\rightarrow0$) under the transformation $a=u\left(  \tau\right)  v^{3}\left(
\tau\right)  $,~$b=v^{3}\left(  \tau\right)  ,~dt=ab^{2}d\tau$ the
gravitational field equations lead to
\begin{equation}
u\left(  \tau\right)  =u_{0}e^{u_{1}\tau},\label{eer.20}%
\end{equation}
while $v\left(  \tau\right)  $ satisfies the two equations%
\begin{equation}
\frac{6u_{1}}{v^{2}}\frac{d}{d\tau}\left(  v^{2}\right)  +\frac{27}{v^{2}%
}\left(  \frac{dv}{d\tau}\right)  ^{2}=-\left(  \rho_{g0}v^{6}e^{\frac{2}%
{3}u_{1}\tau}+\rho_{m0}v^{9}e^{u_{1}\tau}\right)  ,\label{eer.21}%
\end{equation}
and
\begin{equation}
\frac{1}{v^{2}}\left(  \frac{dv}{d\tau}\right)  ^{2}-\frac{1}{\ \nu}%
\frac{d^{2}v}{dt^{2}}+\frac{2}{3}\rho_{g0}v^{6}e^{\frac{2}{3}u_{1}\tau}%
+\rho_{m0}v^{9}e^{u_{1}\tau}=0,\label{eer.22}%
\end{equation}
where we have assumed $\gamma=\frac{4}{3}$for the ghost field. \ When we
perform the coordinate transformation, $v\left(  \tau\right)  =e^{-\frac
{u_{1}}{9}\tau}V\left(  \tau\right)  $, the second-order differential equation
(\ref{eer.22}) is simplified to%
\begin{equation}
\frac{d^{2}V}{d\tau^{2}}=\frac{1}{V}\left(  \frac{dV}{d\tau}\right)
^{2}+\frac{2\rho_{g0}V^{8}+\rho_{m0}V^{11}}{18V}%
\end{equation}
which does not admit any periodic solutions. More specifically, it admits the
unique critical point, for $\rho_{g0}<0$,~$V_{c}=\left(  \frac{2\left\vert
\rho_{g0}\right\vert }{r_{0}}\right)  ^{\frac{1}{3}}$, which is a source point
and describes an Einstein static universe. Now, in the case where $u_{1}=0$,
we have $a\left(  \tau\right)  =v\left(  \tau\right)  $ so the Bianchi I
spacetime reduces to the spatially flat homogeneous FLRW universe.

\begin{figure}[ptb]
\includegraphics[width=5.5cm]{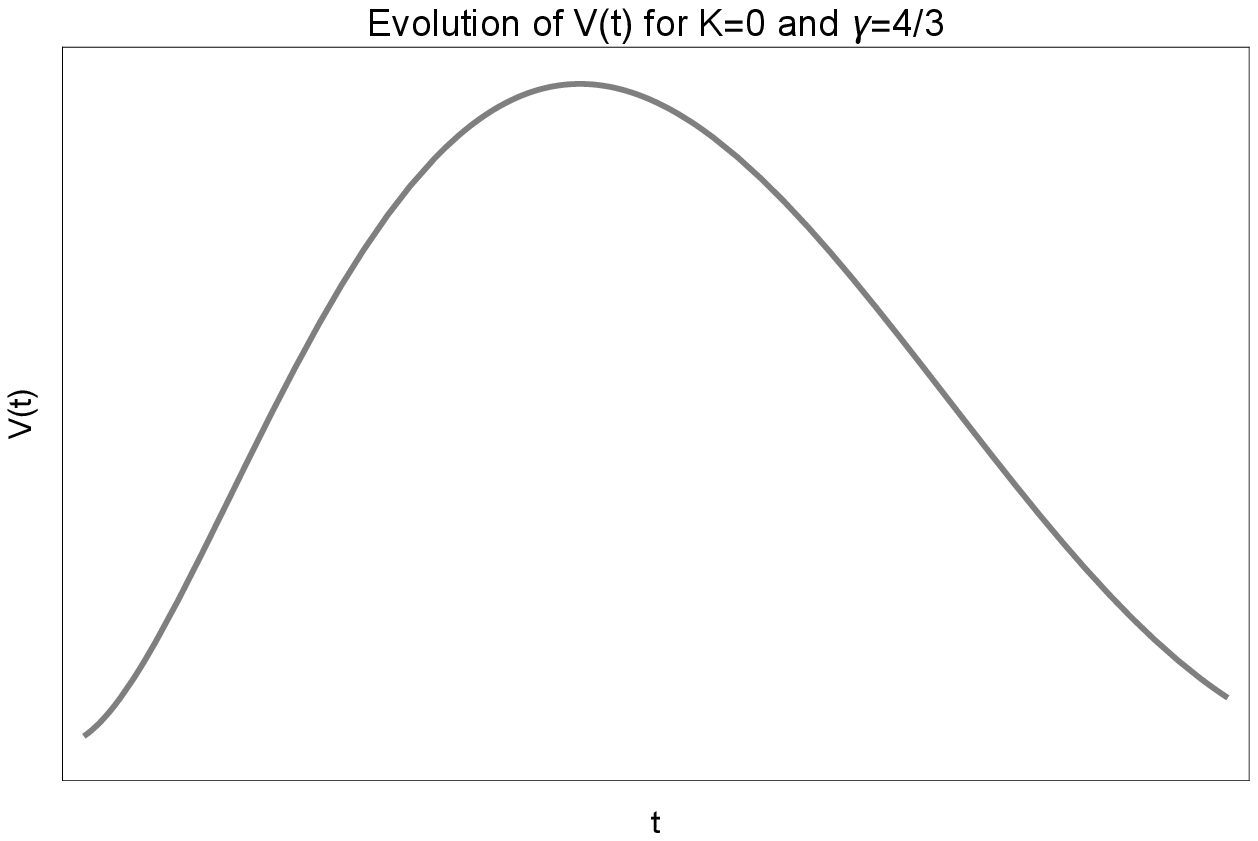}
%\mbox{\epsfxsize=14.2cm \epsffile{numericmodel2q.eps}}
\includegraphics[width=5.5cm]{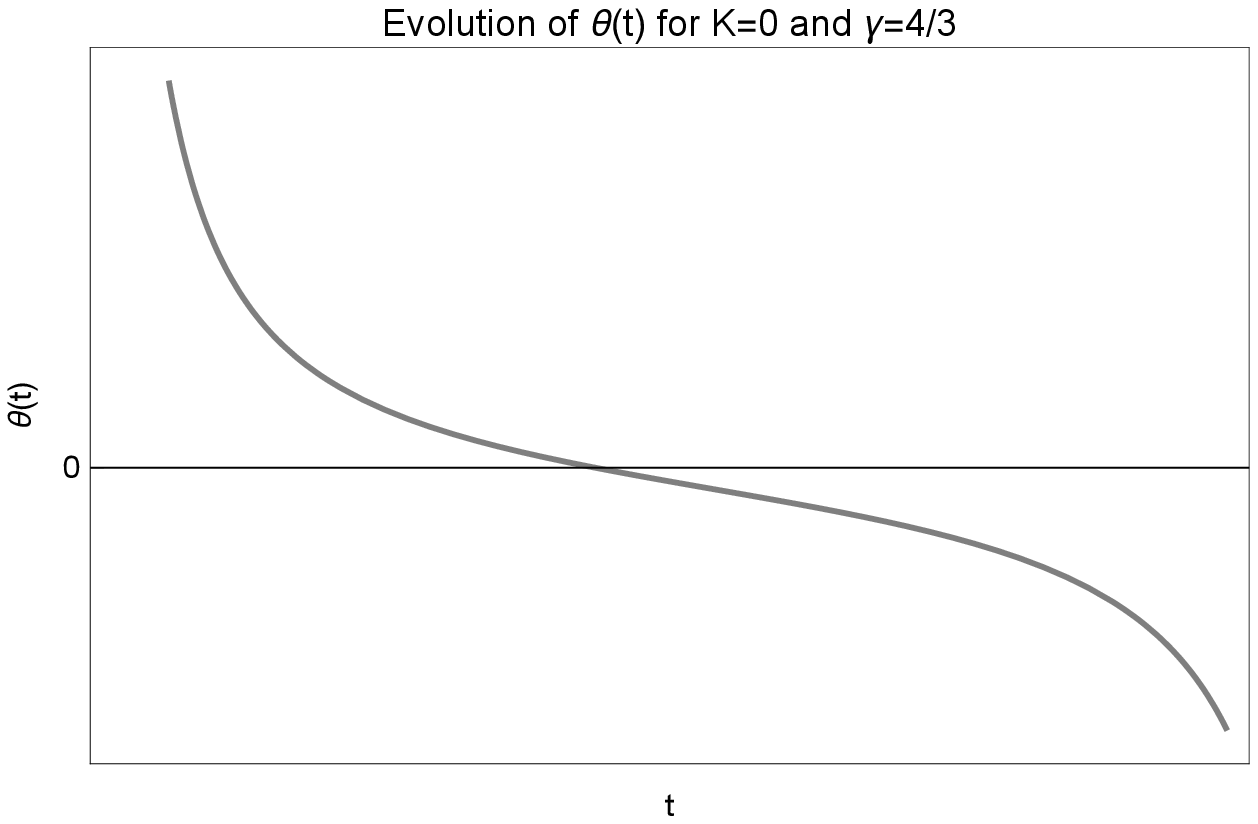}
%\mbox{\epsfxsize=14.2cm \epsffile{numericmodel2q.eps}}
\includegraphics[width=5.5cm]{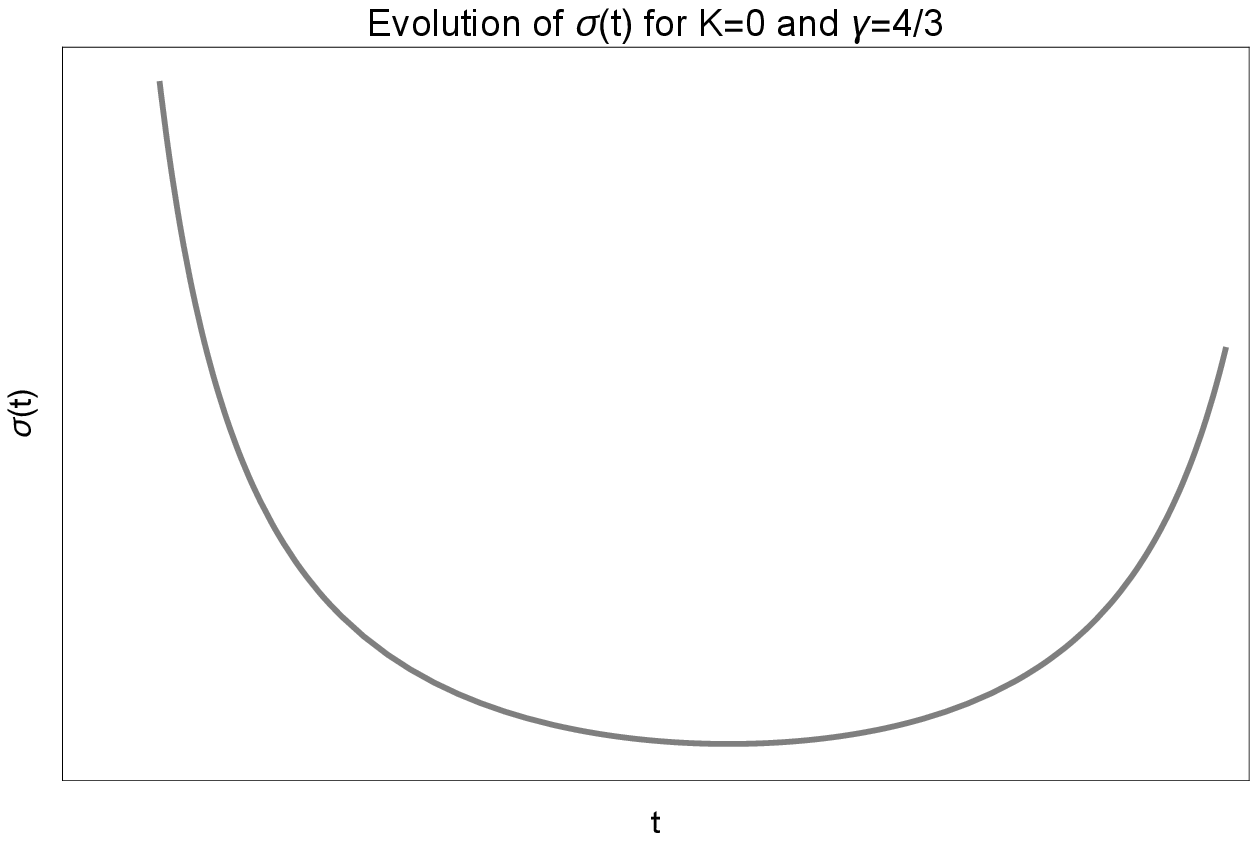}
%\mbox{\epsfxsize=14.2cm \epsffile{numericmodel2q.eps}}
\caption{Numerical simulation of the total volume,~$V\left(  t\right)  $,
volume expansion rate, $\theta\left(  t\right)  $, and shear anisotropy
scalar, $\sigma\left(  t\right)  ,$ for the graviational equations when
$\gamma=\frac{4}{3}$, $\left\vert \rho_{g0}\right\vert >\rho_{m0}$ and $K=0$,
which corresponds to a Bianchi I universe.}%
\label{fig01}%
\end{figure}

In Fig. \ref{fig01} the qualitative time-evolution of the volume $V\left(
t\right)  $, the expansion rate $\theta\left(  t\right)  ,$ and the shear
$\sigma\left(  t\right)  $ are presented following a numerical simulation of
the field equations for $K=0$, and $\left\vert \rho_{g0}\right\vert >\rho
_{m0}$.

In the following section we find that solutions with volume
expansion turning points, where $\theta\left(  t_{0}\right)  =0$ , exist for
$K<0,$ but for different values of the barotropic parameter $\gamma$.

The second family of the Szekeres solutions is that of the inhomogeneous
FLRW-like spacetimes, where the line element is given by the expression
\cite{jdbsc}
\begin{equation}
ds^{2}=-dt^{2}+a^{2}\left(  t\right)  \left(  \left(  \frac{\partial C\left(
r,y,z\right)  }{\partial r}\right)  ^{2}dr^{2}+e^{2C\left(  r,y,z\right)
}\left(  dy^{2}+dz^{2}\right)  \right)  . \label{sf.15}%
\end{equation}
The spatial function $C\left(  r,y,z\right)  ~$is given by the expression%
\begin{equation}
C\left(  r,y,z\right)  =-\ln\left(  \gamma_{1}\left(  r\right)  \left(
\left(  y-\gamma_{2}\left(  r\right)  \right)  ^{2}+\left(  z-\gamma
_{3}\left(  r\right)  \right)  ^{2}\right)  +\gamma_{4}\left(  r\right)
\right)  , \label{sf.16}%
\end{equation}
where two of the four arbitrary functions, $\gamma_{1}\left(  r\right)
\rightarrow\gamma_{4}\left(  r\right)  ,$ are related to the spatial
curvature, $K$, by%
\begin{equation}
K=8\gamma_{1}\left(  r\right)  \gamma_{4}\left(  r\right)  . \label{sf.17}%
\end{equation}
It is important to mention here that $K$ is a constant and not a function of
$r$ as it is in the case of the Szekeres spacetimes. This difference arises
because of the existence of the second (homogeneous ghost) fluid source.
Moreover, the evolution of scale factor, $a\left(  t\right)  ,$ is described
by Friedmann's equations with two homogeneous perfect fluids; its general
analytic solution is expressed in terms of elliptic integrals.

However, in the particular case for which $\rho_{g}$ describes a radiation
ghost field, i.e., $\gamma=\frac{4}{3}$, the exact form of the scale factor is
given by the following simple expression \cite{jts, jkm}%
\begin{equation}
a\left(  \tau\right)  =\frac{\rho_{m0}}{6k}+\sqrt{\left(  \frac{\rho_{m0}}%
{6k}\right)  ^{2}-\frac{\left\vert \rho_{g0}\right\vert }{3k}}\sin\left(
\sqrt{K}\tau\right)  ~\text{for}~K\neq0, \label{ee.20}%
\end{equation}
or by
\begin{equation}
a\left(  \tau\right)  =\frac{\left\vert \rho_{g0}\right\vert }{\rho_{m0}%
}+\frac{\rho_{m0}}{12}\tau^{2}~\text{for }K=0, \label{ee.21}%
\end{equation}
where $\tau$ is the conformal time defined by $dt=a\left(  \tau\right)  d\tau
$. The scale factor in $K=0$ solution increases towards a power law, with a
minimum as $\tau\rightarrow0$ at $a\left(  0\right)  =$ $\frac{\left\vert
\rho_{g0}\right\vert }{\rho_{m0}}$. For zero spatial curvature the scale
factor has a minimum at $a_{\min}^{\left(  k=0\right)  }=\frac{\left\vert
\rho_{g0}\right\vert }{\rho_{m0}}$. For positive spatial curvature, ($K=1$),
the solution (\ref{ee.20}) is real when $\left(  \rho_{m0}\right)
^{2}>12\left\vert \rho_{g0}\right\vert $ and it is also a periodic solution
with minimum and maximum of $a(\tau$) and $a(t)$ at%
\begin{equation}
a_{\min}^{\left(  k=1\right)  }=\frac{1}{6}\left(  \rho_{m0}-\sqrt{\rho
_{m0}^{2}-12\left\vert \rho_{g0}\right\vert }\right)  ~,~a_{\max}^{\left(
k=1\right)  }=\frac{1}{6}\left(  \rho_{m0}+\sqrt{\rho_{m0}^{2}-12\left\vert
\rho_{g0}\right\vert }\right)  , \label{ee.22}%
\end{equation}
and the scale factor can be written as \cite{jts}%
\begin{equation}
a^{\left(  k=1\right)  }\left(  \tau\right)  =\frac{1}{2}\left[  \left(
a_{\max}^{\left(  k=1\right)  }+a_{\min}^{\left(  k=1\right)  }\right)
+\left(  a_{\max}^{\left(  k=1\right)  }-a_{\min}^{\left(  k=1\right)
}\right)  \sin\left(  \tau\right)  \right]  \label{ee.22aa}%
\end{equation}
so we can see that the scale factor oscillates around the static solution
$a^{\left(  k=1\right)  }\left(  0\right)  =\left(  a_{\max}^{\left(
k=1\right)  }+a_{\min}^{\left(  k=1\right)  }\right)  $ with arbitrary
amplitude. Hence these solutions show the stability of the Einstein static
universe to these bounded oscillations but they only occur when a ghost field
is present.

Note that the quantities $a_{\max}^{\left(  k=1\right)  }$ and $a_{\min
}^{\left(  k=1\right)  }$ are not spatially varying because they depend on the
constant quantities $\rho_{m0}$ and $\left\vert \rho_{g0}\right\vert $.

Finally, for $K=-1$, solution (\ref{ee.20}) is real if and only if $\rho
_{m0}<12\left\vert \rho_{g0}\right\vert $ and the scale factor then simplifies
to%
\begin{equation}
a\left(  \tau\right)  =\frac{\rho_{m0}}{6}+\sqrt{\frac{\left\vert \rho
_{g0}\right\vert }{3}-\left(  \frac{\rho_{m0}}{6}\right)  ^{2}}\sinh\left(
\tau\right)  , \label{ee.23}%
\end{equation}
which increases exponentially as $\tau\rightarrow\infty$.

We continue our analysis by studying the stability of these particular
solutions with emphasis on the cyclic solutions.

\section{Stability of the cyclic solutions}

\label{cyc1}

We have seen that the addition of the ghost field to the Szekeres universes
can create new cyclic solutions, or solutions in which the volume expansion
rate, $\theta\left(  t\right)  ,$ can go to zero and change sign. In this
section we perform a dynamical analysis of the kinematic quantities for the
gravitational field equations. Here, the solutions with $\theta=0$ appear and
we are able to study their stability.

In terms of the kinematic quantities $\theta,~\sigma,~\mathcal{E}$,~$\rho_{m}$
and\footnote{Here, $\sigma$ denotes the shear scalar and $\mathcal{E}$~is the
scalar for the electric part of the Weyl tensor.} $\rho_{g},$ the Szekeres
field equations (\ref{ee.02}) are expressed as follows \cite{ellis1,ellis2}%

\begin{align}
\frac{d\rho_{m}}{dt}+\theta\rho &  =0,\label{ss.01}\\
~\frac{d\rho_{g}}{dt}+\gamma\rho_{g}\theta &  =0\label{ss.01a}\\
\frac{d\theta}{dt}+\frac{\theta^{2}}{3}+6\sigma^{2}+\frac{1}{2}\rho_{m}%
+\frac{\left(  3\gamma-2\right)  }{2}\rho_{\gamma}  &  =0,\label{ss.02}\\
\frac{d\sigma}{dt}-\sigma^{2}+\frac{2}{3}\theta\sigma+\mathcal{E}  &
=0,\label{ss.04}\\
\frac{d\mathcal{E}}{dt}+3\mathcal{E}\sigma+\theta\mathcal{E}+\left(  \frac
{1}{2}\rho_{m}+\frac{\gamma}{2}\rho_{g}\right)  \sigma &  =0, \label{ss.05}%
\end{align}%
\begin{equation}
\frac{\theta^{2}}{3}-3\sigma^{2}+\frac{^{\left(  3\right)  }R}{2}-\rho
_{m}-\rho_{g}=0, \label{ss.06}%
\end{equation}
where $^{\left(  3\right)  }R$ denotes the curvature of the three-dimensional hypersurfaces.

We proceed by choosing the new dimensionless variables \cite{alg}, $\omega
_{m},\omega_{r}$ and $\omega_{R}$ defined via
\begin{equation}
\rho_{m}=\frac{1}{3}\omega_{m}\left(  1+\theta^{2}\right)  ~,~\rho_{r}%
=\frac{1}{3}\omega_{r}\left(  1+\theta^{2}\right)  ~,~^{\left(  3\right)
}R=\frac{2}{3}\omega_{R}\left(  1+\theta^{2}\right)  \label{ss.07a}%
\end{equation}
and $\beta,\mathcal{E}$ and $h$ by
\begin{equation}
~\sigma=\frac{1}{\sqrt{3}}\beta\sqrt{1+\theta^{2}}~,~\varepsilon=\frac{1}%
{3}\mathcal{E}\left(  1+\theta^{2}\right)  ~,~~h^{2}=\left(  \frac{\theta
}{\sqrt{1+\theta^{2}}}\right)  ^{2}, \label{ss.07b}%
\end{equation}
so the gravitational field equations become an autonomous system:%
\begin{equation}
\frac{d\omega_{m}}{d\zeta}=\frac{1}{3}h\omega_{m}\left(  2h^{2}+12\beta
^{2}+\omega_{m}+2\omega_{g}-3\right)  , \label{se.01}%
\end{equation}%
\begin{equation}
\frac{d\omega_{r}}{d\zeta}=\frac{1}{3}h\omega_{r}\left(  2h^{2}+12\beta
^{2}+\omega_{m}+2\omega_{g}-3\gamma\right)  , \label{se.02}%
\end{equation}%
\begin{equation}
\frac{d\beta}{d\zeta}=\frac{1}{6\sqrt{3}}\left[  \beta\left(  6\beta+\sqrt
{3}h\left(  2h^{2}-4+12\beta^{2}+\omega_{m}+2\omega_{g}\right)  \right)
-6\varepsilon\right]  , \label{se.03}%
\end{equation}%
\begin{equation}
\frac{d\varepsilon}{d\zeta}=\frac{1}{6}\left[  4h^{3}\varepsilon
+2h\varepsilon\left(  12\beta^{2}+\omega_{m}+2\omega_{g}-3\right)  -\sqrt
{3}\beta\left(  6\varepsilon+\omega_{m}+\omega_{g}\right)  \right]  ,
\label{se.04}%
\end{equation}%
\begin{equation}
\frac{dh}{d\zeta}=\frac{1}{6}\left(  h^{2}-1\right)  \left(  2h^{2}%
+12\beta^{2}+\omega_{m}+2\omega_{g}\right)  , \label{se.05}%
\end{equation}
and there is a first integral%
\begin{equation}
\omega_{R}=\omega_{m}+\omega_{g}+3\beta^{2}-h^{2}, \label{se.06}%
\end{equation}
where the new time variable, $\zeta$, is defined as $dt=\left(  \sqrt
{1+\theta^{2}}\right)  d\zeta.$

This normalization of the variables differs from the usual $H-$normalization
\cite{ellis1, cope} because now it is possible to determine critical points
also in the surface where $\theta=0$, where $h=0$. Furthermore, parameters
$\omega_{m}$,~$\omega_{g}$ and $\omega_{R}$ are related to the familiar
energy-density parameters~$\Omega_{m},~\Omega_{g}$ and $\Omega_{R}$ as
follows:%
\begin{equation}
\omega_{m}=\Omega_{m}h^{2}~,~\omega_{g}=\Omega_{g}h^{2}~~\text{and~ }%
\omega_{r}=\Omega_{R~}h^{2}~. \label{ss.08b}%
\end{equation}

We are interested in the critical points for the system (\ref{se.01}%
)-(\ref{se.05}) when $\theta$ is zero. They can be easily computed:%
\begin{equation}
P_{1}:\left(  h,\beta,\varepsilon,\omega_{m},\omega_{g}\right)  =\left(
0,\beta,\beta^{2},12\frac{\left(  \gamma-1\right)  }{2-\gamma}\beta^{2}%
,-\frac{6\beta^{2}}{2-\gamma}\right)
\end{equation}
and
\begin{equation}
P_{2}:\left(  h,\beta,\varepsilon,\omega_{m},\omega_{g}\right)  =\left(
0,0,0,\omega_{m},-\frac{\omega_{m}}{2}\right)  .
\end{equation}
These points $P_{1}~$and $P_{2}$ describe Einstein static universes.

However, in addition to those two critical points there is a family of
critical points where $h^{2}=1$. These correspond to the Szekeres universes
when $\omega_{g}=0$ and to the Szekeres-Szafron universes \cite{szafron} when
$\omega_{m}=0$. Moreover, we find that there is no critical point where
$\omega_{m}\omega_{g}\neq0$. \ 

\begin{figure}[ptb]
\includegraphics[width=5.5cm]{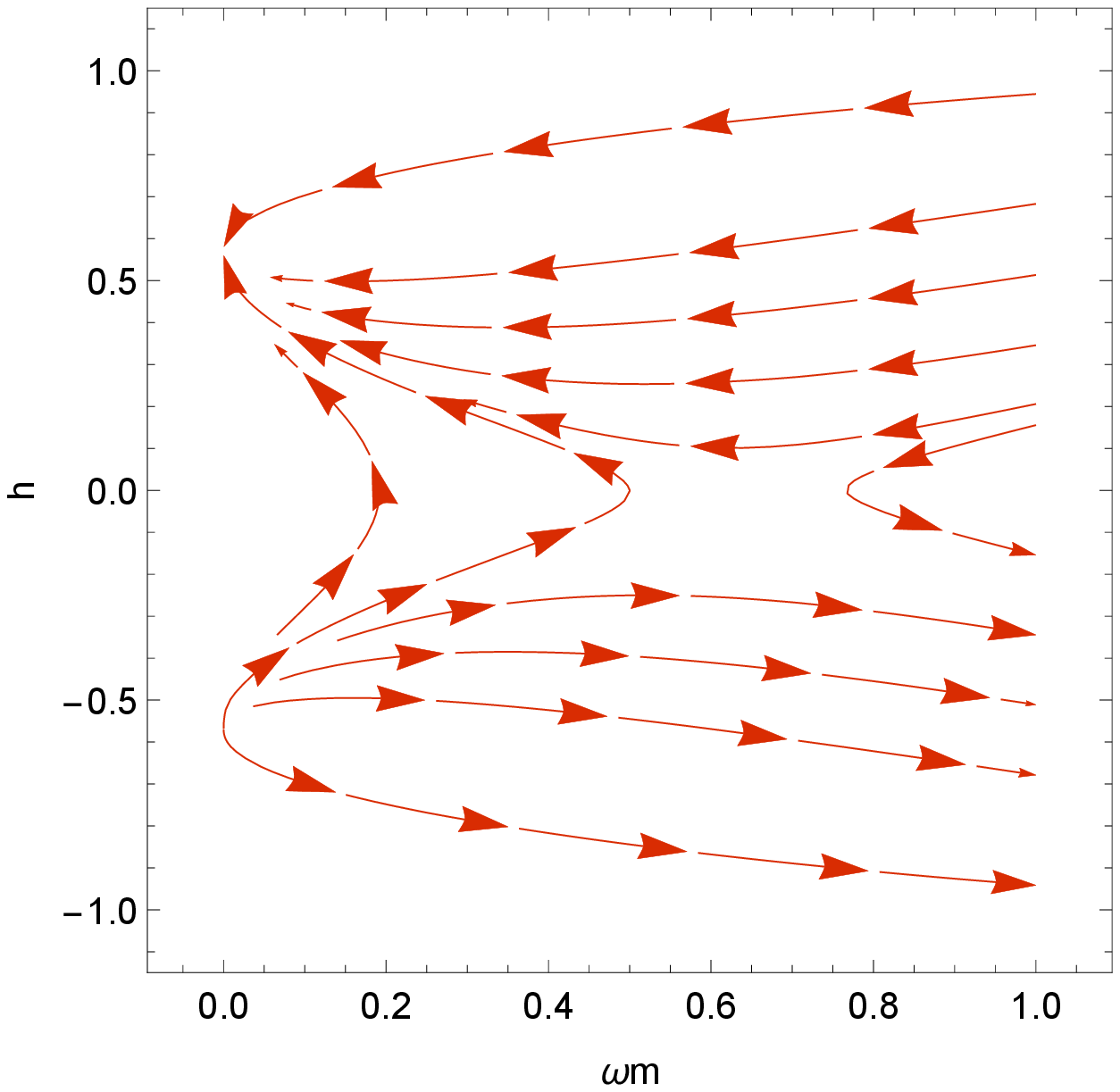}
%\mbox{\epsfxsize=14.2cm \epsffile{numericmodel2q.eps}}
\includegraphics[width=5.5cm]{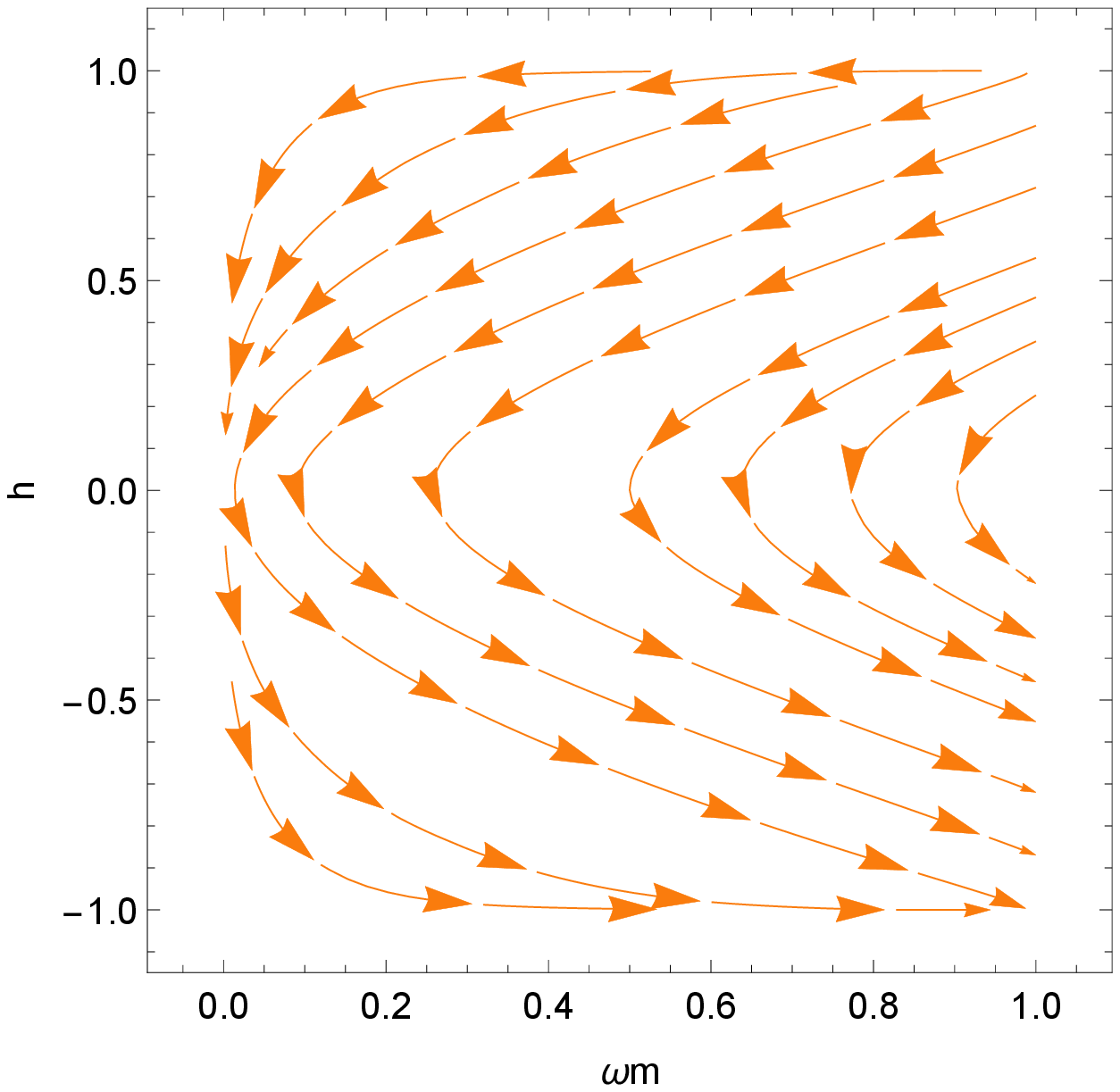}
%\mbox{\epsfxsize=14.2cm \epsffile{numericmodel2q.eps}}
\includegraphics[width=5.5cm]{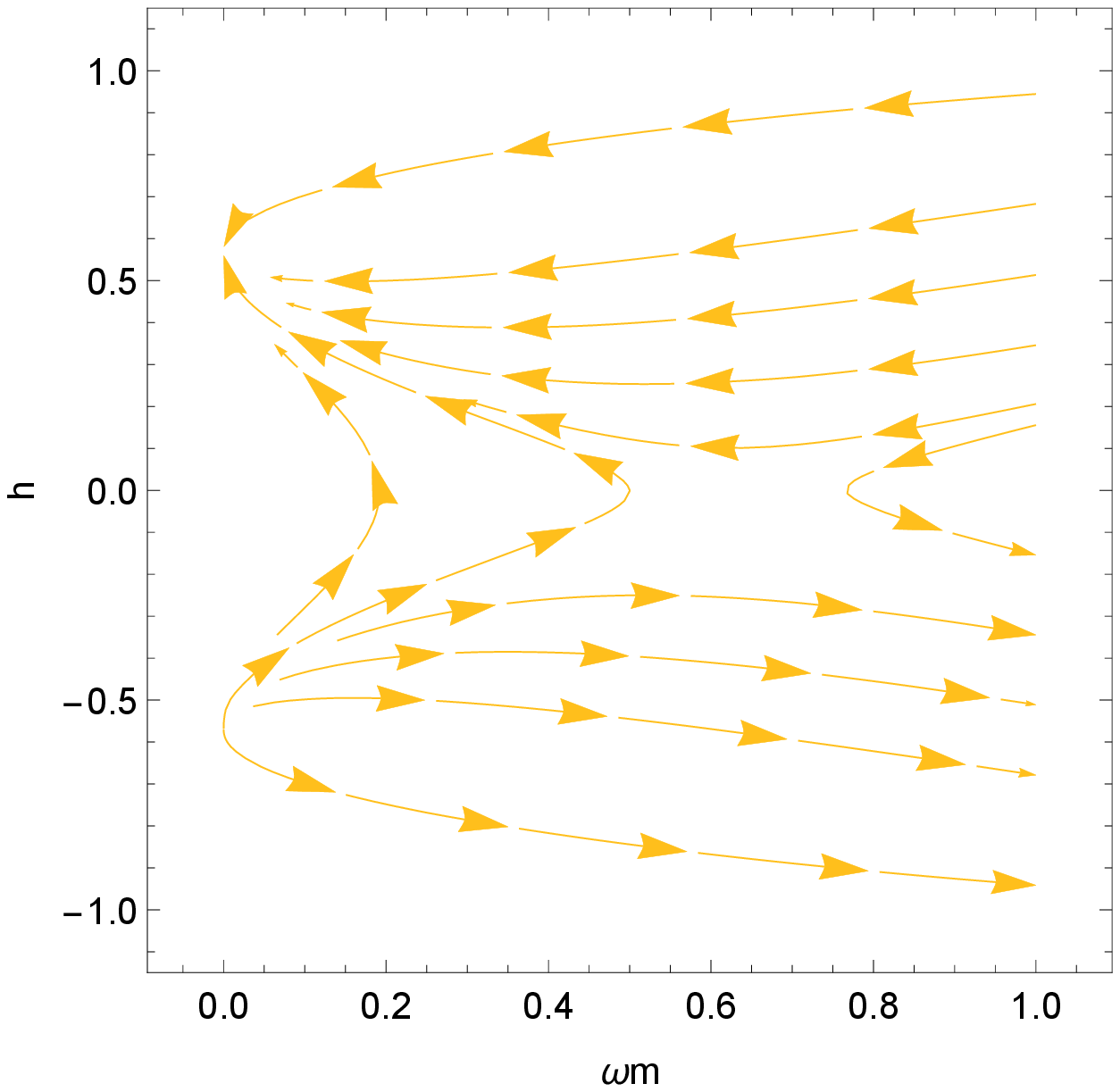}
%\mbox{\epsfxsize=14.2cm \epsffile{numericmodel2q.eps}}
\caption{Phase-space diagram for the dynamical system (\ref{se.01}%
)-(\ref{se.05}) in the $\omega_{m}-h$ surface and for three different values
of $\beta$,~$\gamma=\frac{4}{3}$ and $\omega_{g},$ as given by the point
$P_{1}$ The middle figure is for $\beta=0$, the left figure for $\beta<0$ and
the right figure for $\beta>0$.}%
\label{pp01}%
\end{figure}

\begin{figure}[ptb]
\includegraphics[width=5.5cm]{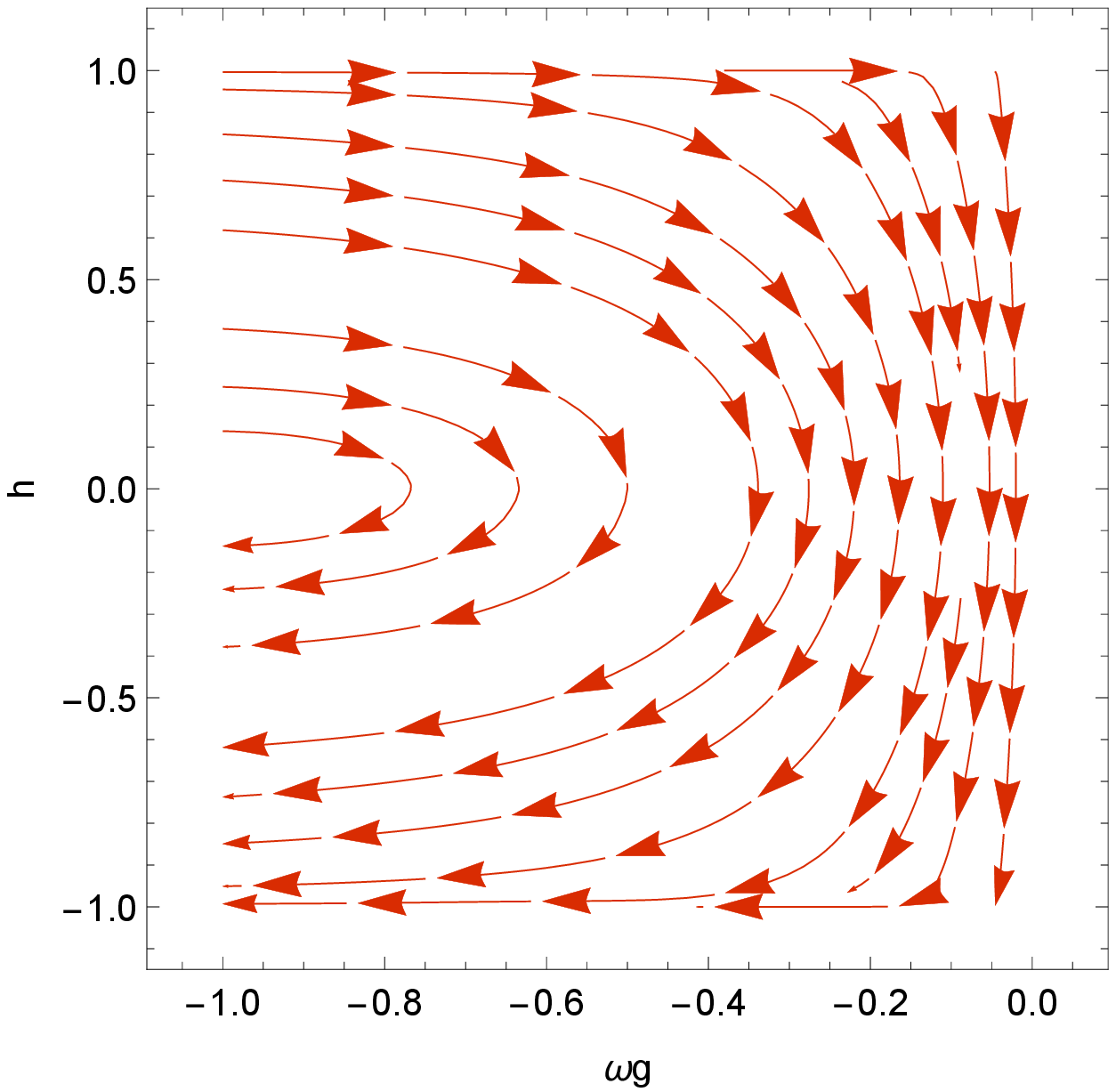}
%\mbox{\epsfxsize=14.2cm \epsffile{numericmodel2q.eps}}
\includegraphics[width=5.5cm]{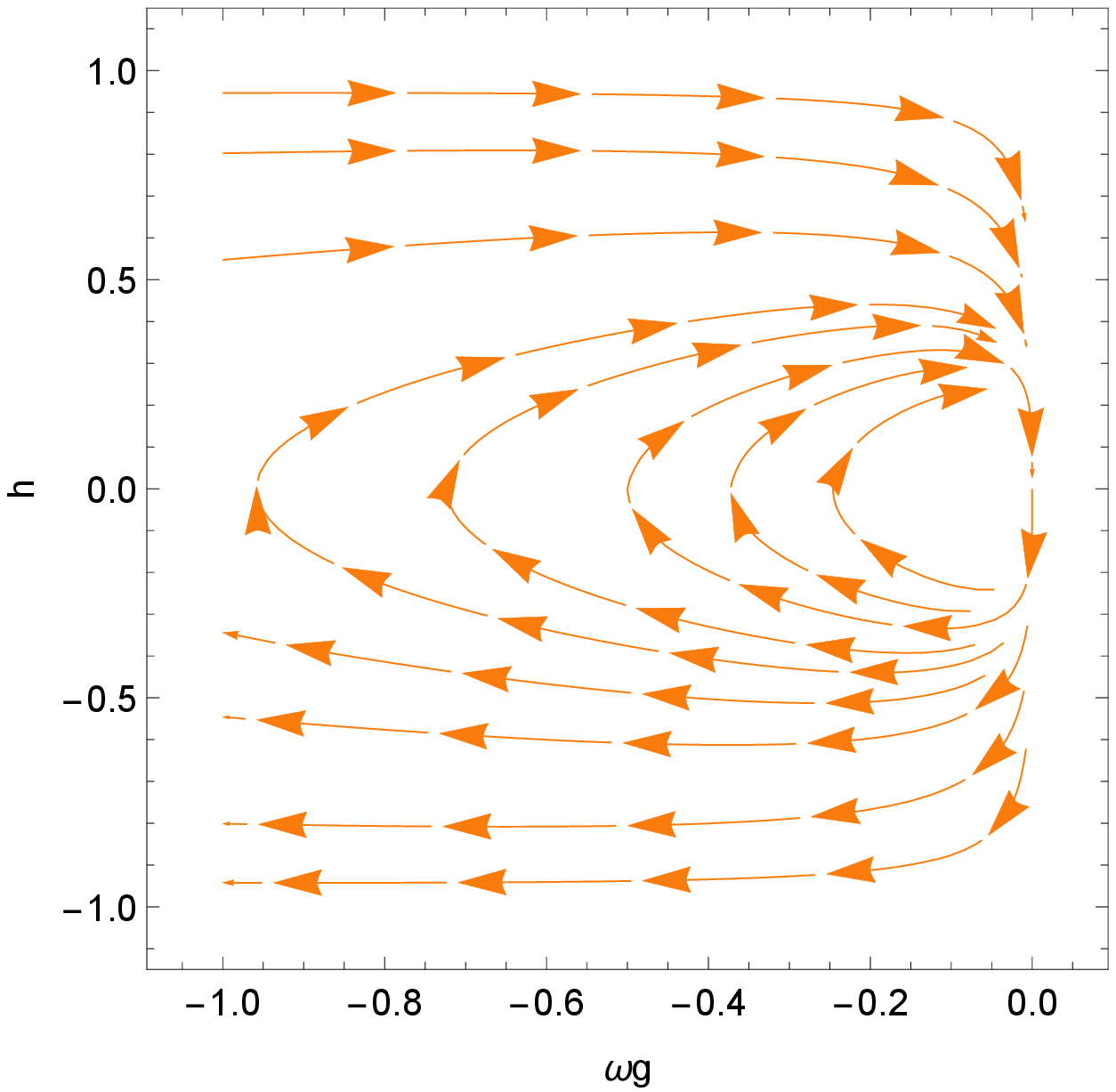}
%\mbox{\epsfxsize=14.2cm \epsffile{numericmodel2q.eps}}
\includegraphics[width=5.5cm]{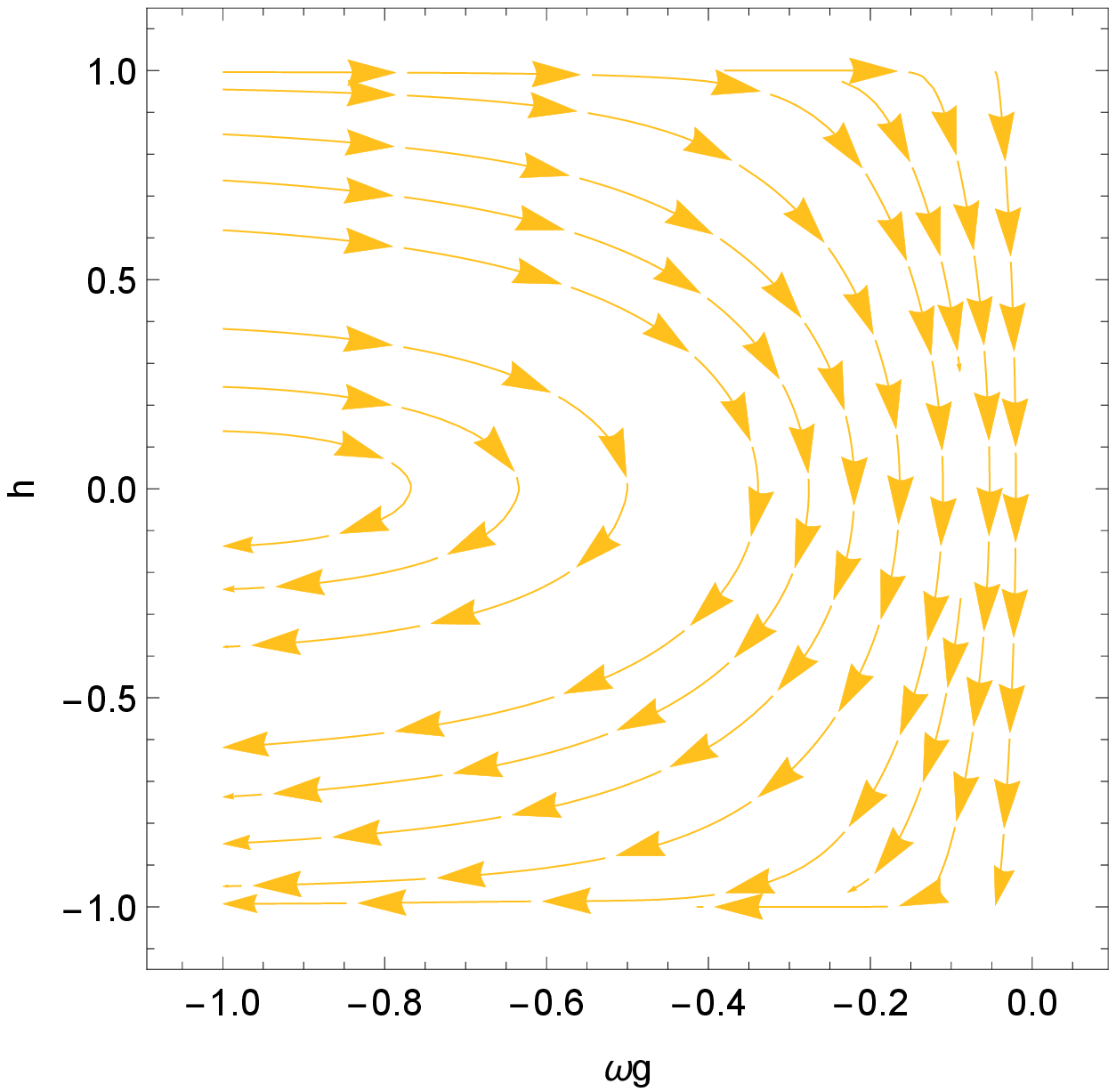}
%\mbox{\epsfxsize=14.2cm \epsffile{numericmodel2q.eps}}
\caption{Phase-space diagram for the dynamical system (\ref{se.01}%
)-(\ref{se.05}) on the $\omega_{g}-h$ surface, ~for $\gamma=\frac{4}{3}$,
$\omega_{m}=12\frac{\left(  \gamma-1\right)  }{2-\gamma}\beta^{2}$ and for
three different values of $\beta$. The middle figure is for $\beta=0$, the
left figure for $\beta<0$ and the right figure for $\beta>0.$}%
\label{pp02}%
\end{figure}

\begin{figure}[ptb]
\includegraphics[width=5.5cm]{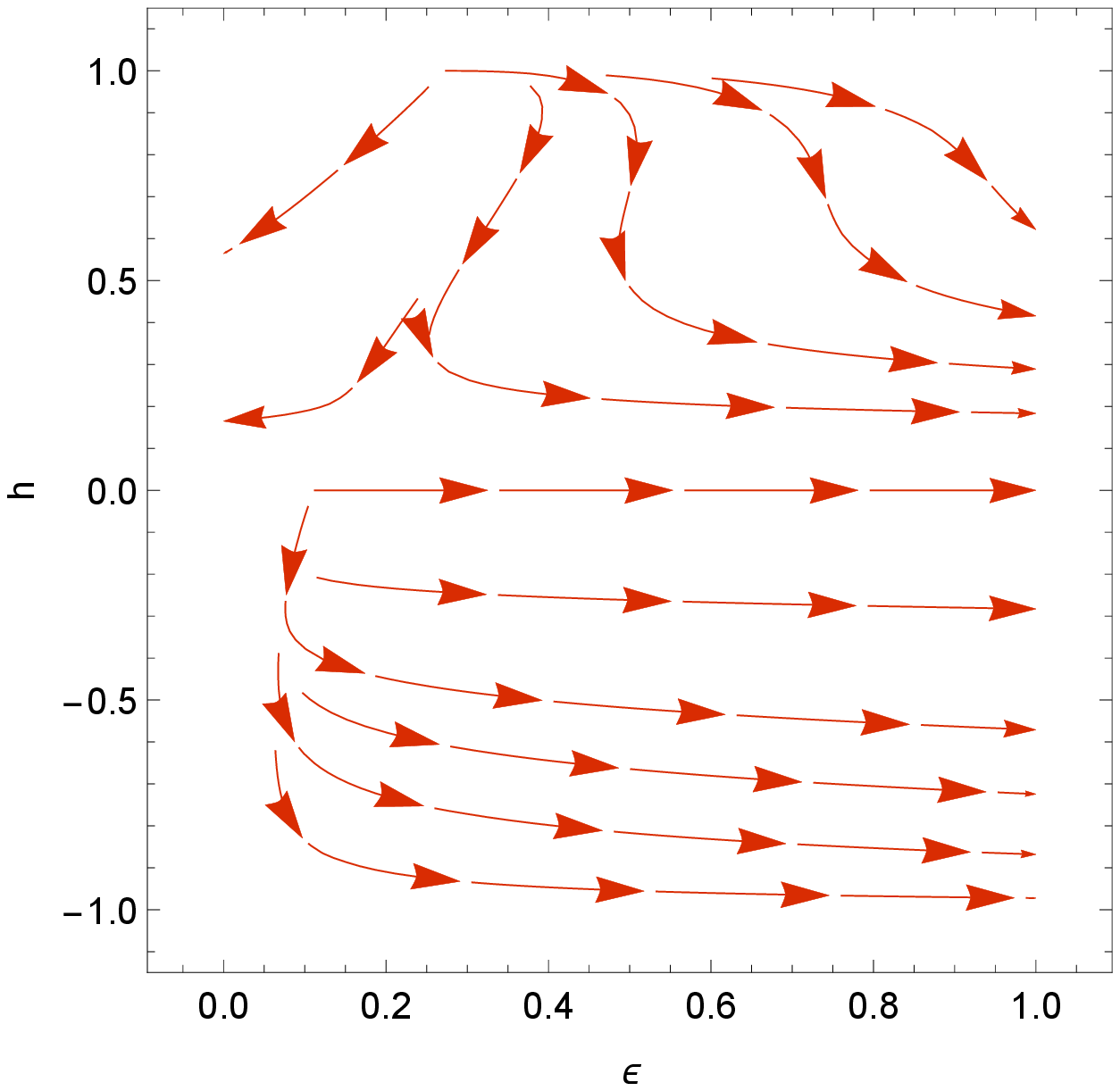}
%\mbox{\epsfxsize=14.2cm \epsffile{numericmodel2q.eps}}
\includegraphics[width=5.5cm]{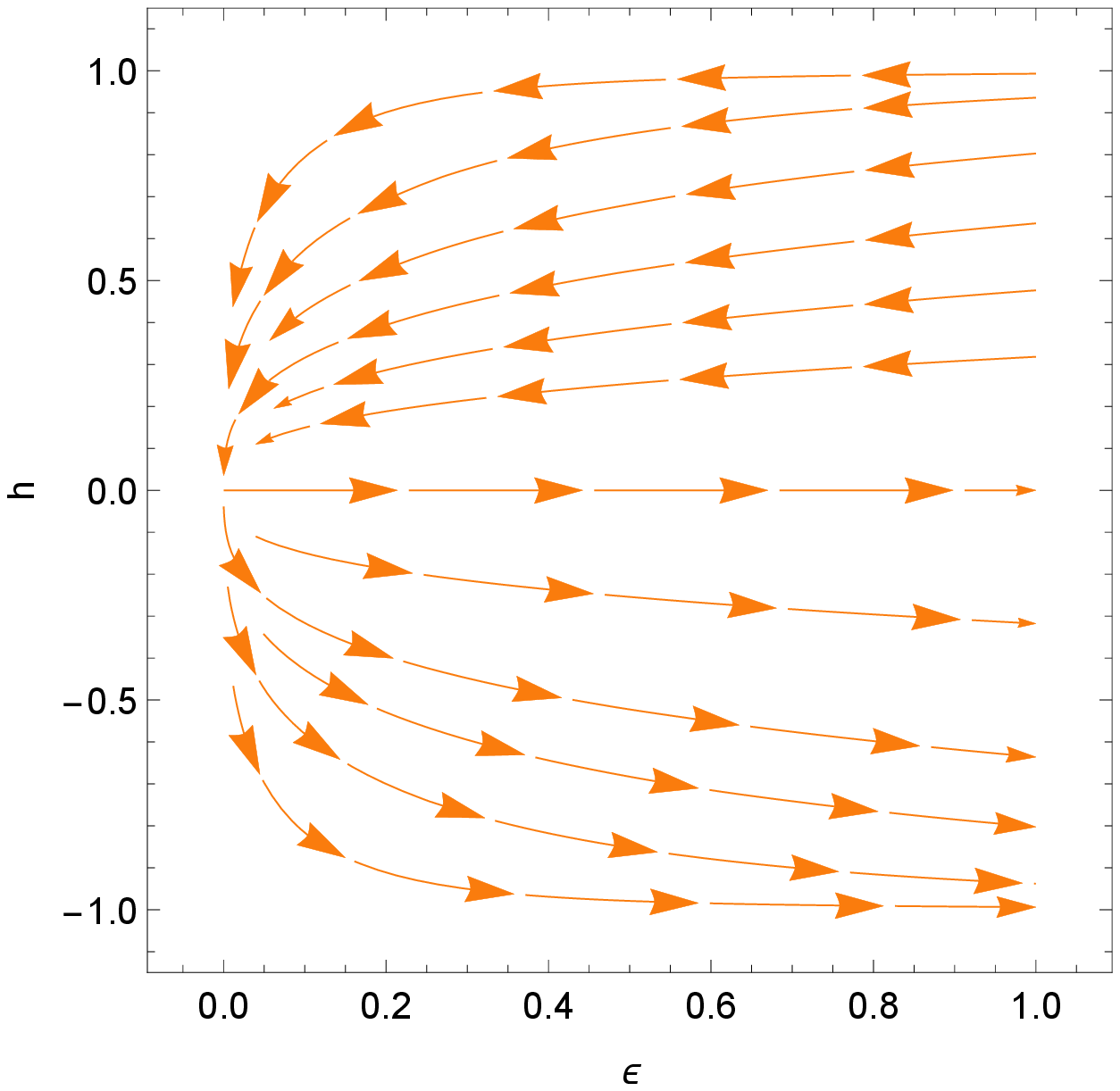}
%\mbox{\epsfxsize=14.2cm \epsffile{numericmodel2q.eps}}
\includegraphics[width=5.5cm]{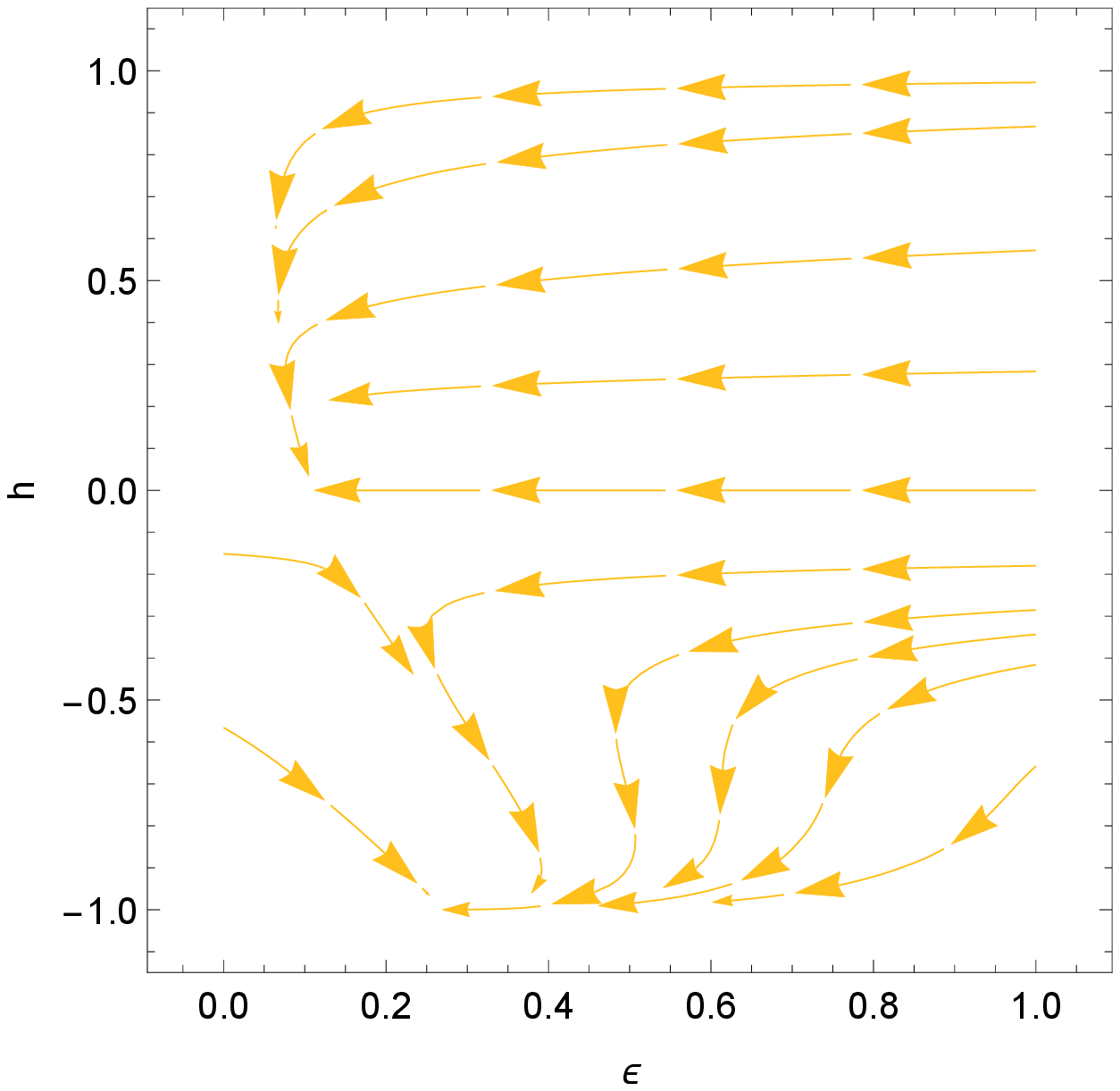}
%\mbox{\epsfxsize=14.2cm \epsffile{numericmodel2q.eps}}
\caption{Phase-space diagram for the dynamical system (\ref{se.01}%
)-(\ref{se.05}) on the $\varepsilon-h$ surface, for~$\gamma=\frac{4}{3}$,~
$\left(  \omega_{m},\omega_{g}\right)  =\left(  12\frac{\left(  \gamma
-1\right)  }{2-\gamma}\beta^{2},-\frac{6\beta^{2}}{2-\gamma}\right)  ~$and for
three different values of $\beta$. The middle figure is for $\beta=0$, the
left figure for $\beta<0$ and the right figure for $\beta>0.$}%
\label{pp03}%
\end{figure}

\begin{figure}[ptb]
\includegraphics[width=5.5cm]{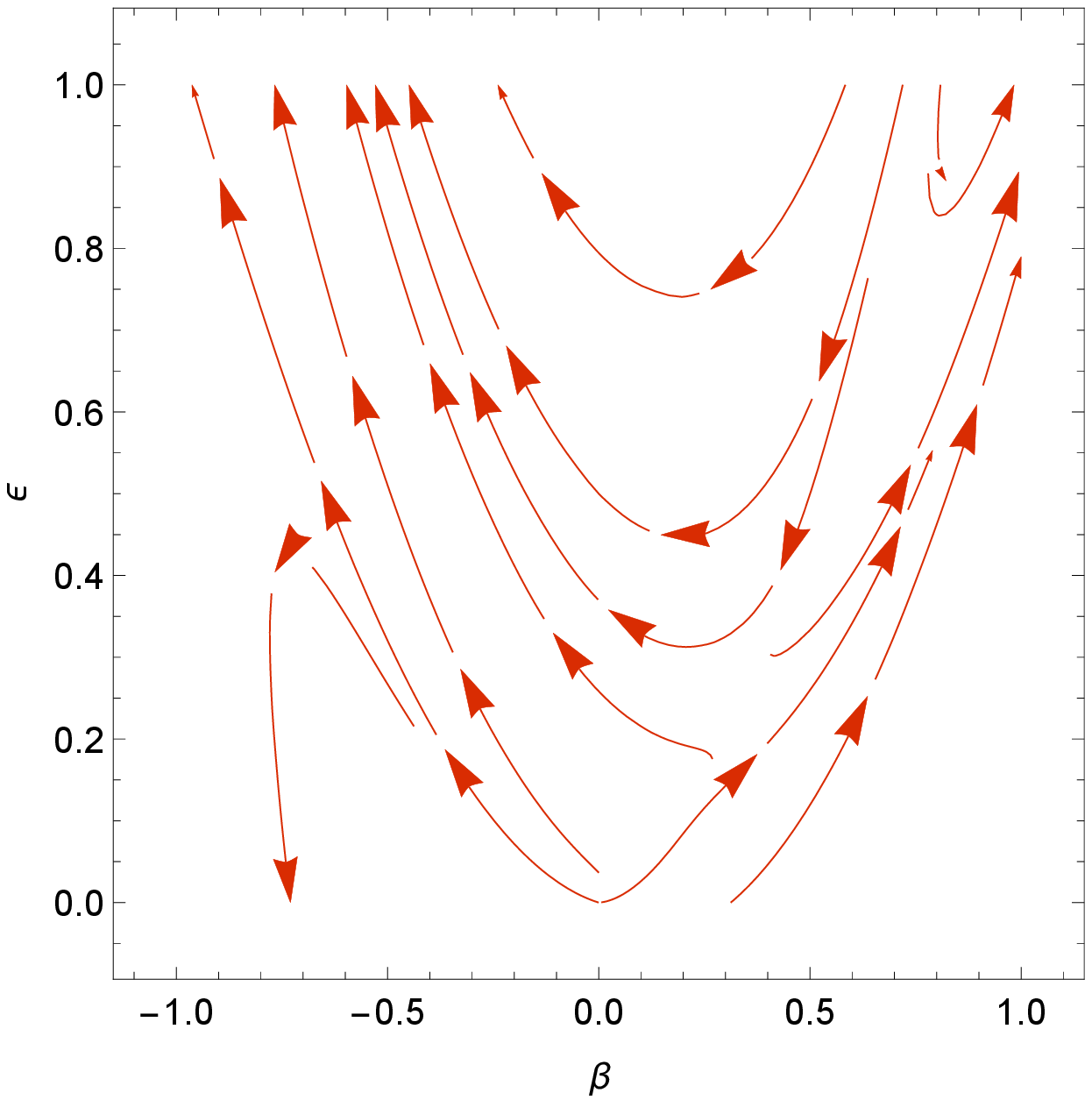}
%\mbox{\epsfxsize=14.2cm \epsffile{numericmodel2q.eps}}
\includegraphics[width=5.5cm]{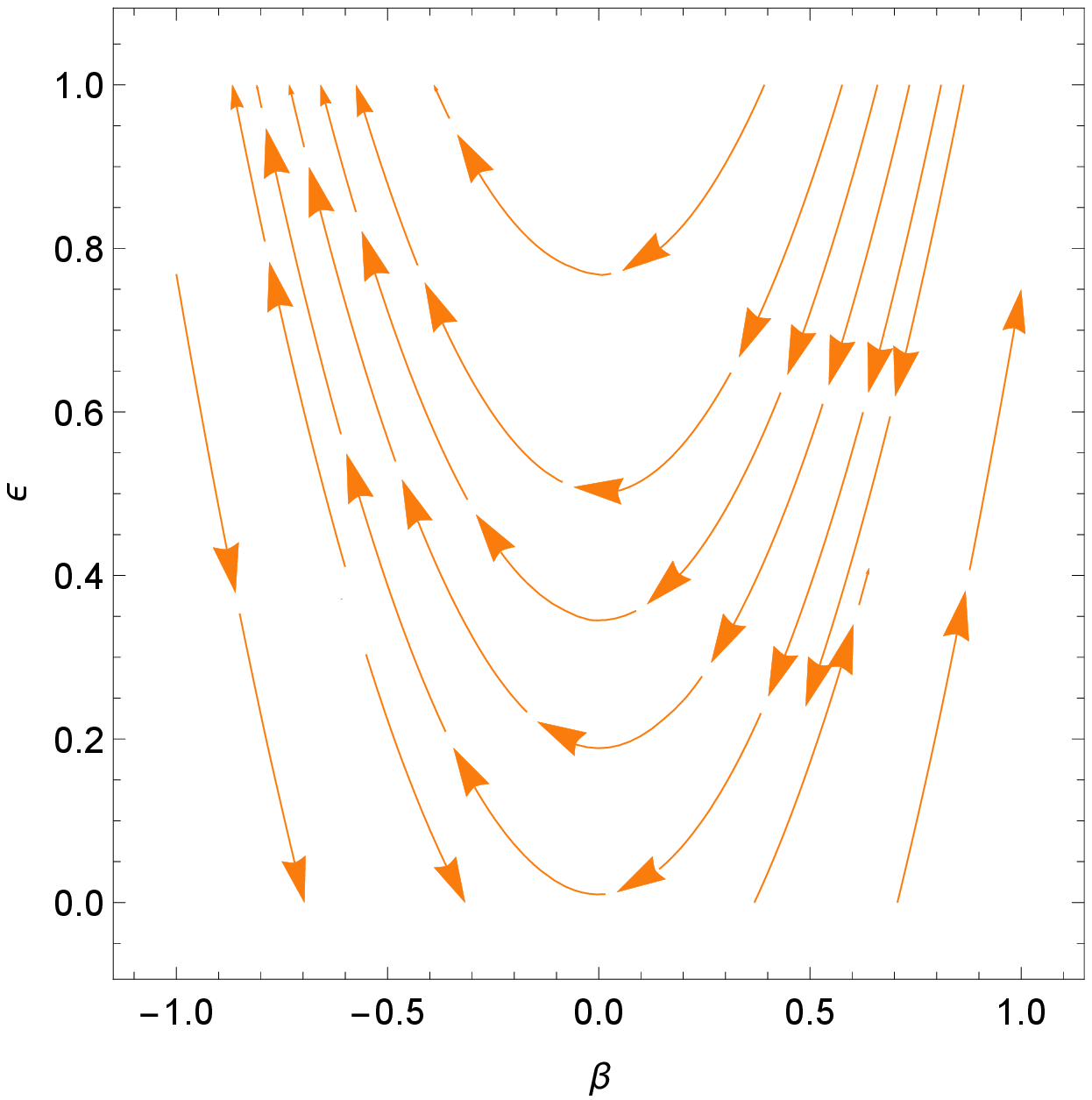}
%\mbox{\epsfxsize=14.2cm \epsffile{numericmodel2q.eps}}
\includegraphics[width=5.5cm]{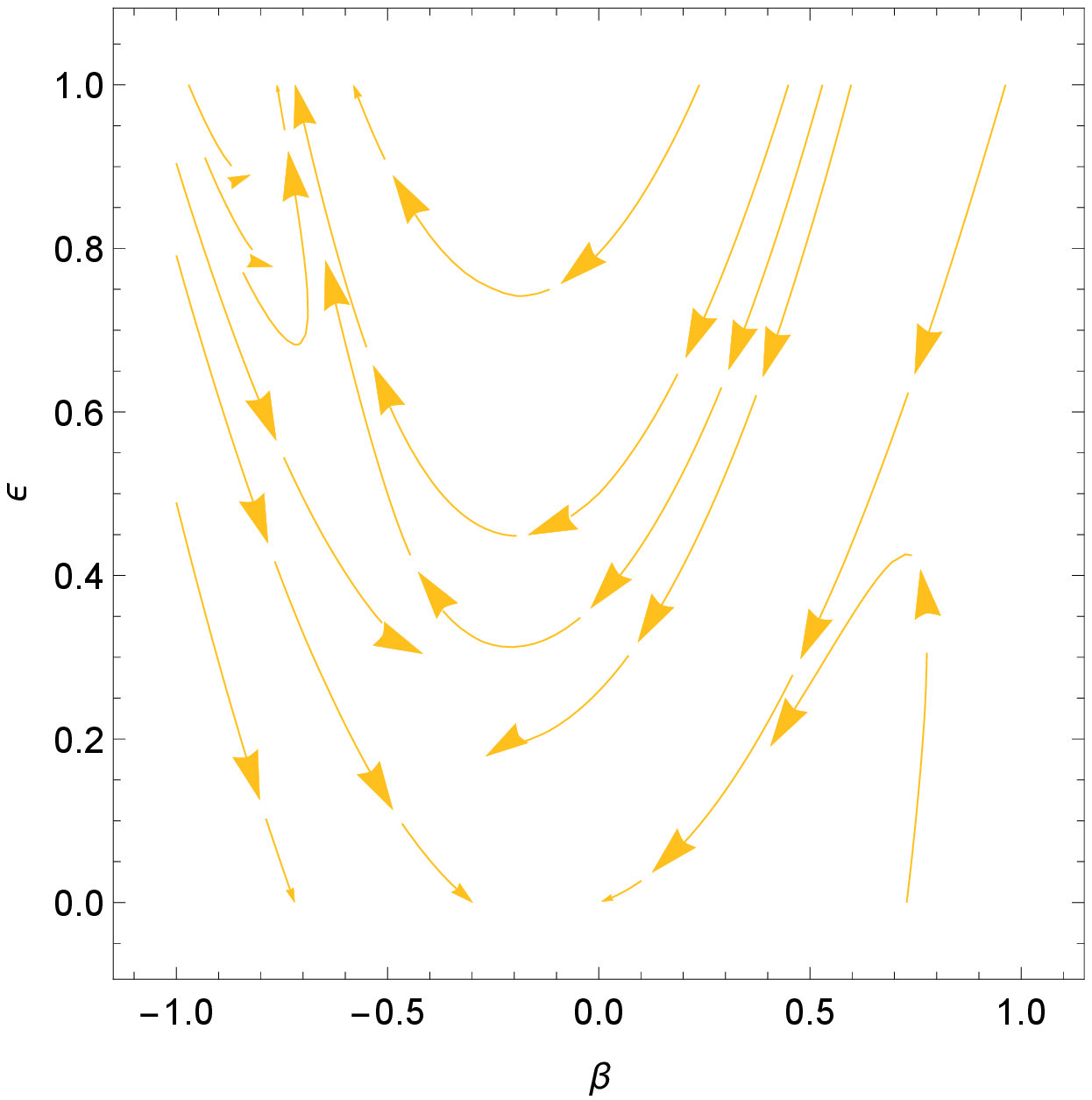}
%\mbox{\epsfxsize=14.2cm \epsffile{numericmodel2q.eps}}
\caption{Phase-space diagram for the dynamical system (\ref{se.01}%
)-(\ref{se.05}) on the $\beta-\varepsilon$ surface$,$ for~$\gamma=\frac{4}{3}%
$,~ $\left(  \omega_{m},\omega_{g}\right)  =\left(  12\frac{\left(
\gamma-1\right)  }{2-\gamma}\beta^{2},-\frac{6\beta^{2}}{2-\gamma}\right)  $
and for three different values of $h$ around the $h\left(  P_{1}\right)  =0$
value. }%
\label{pp04}%
\end{figure}

\begin{figure}[ptb]
\includegraphics[width=5.5cm]{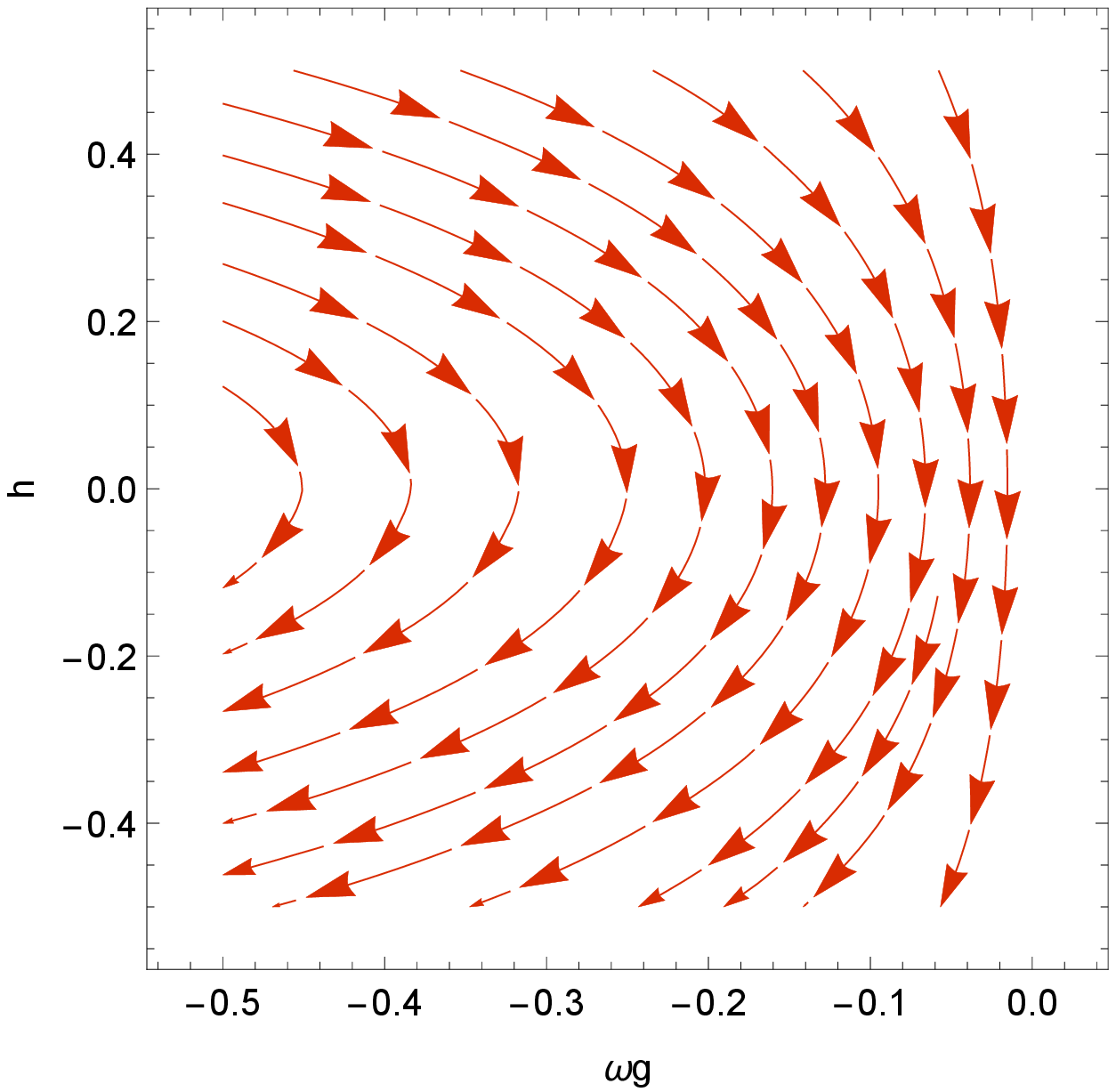}
%\mbox{\epsfxsize=14.2cm \epsffile{numericmodel2q.eps}}
\includegraphics[width=5.5cm]{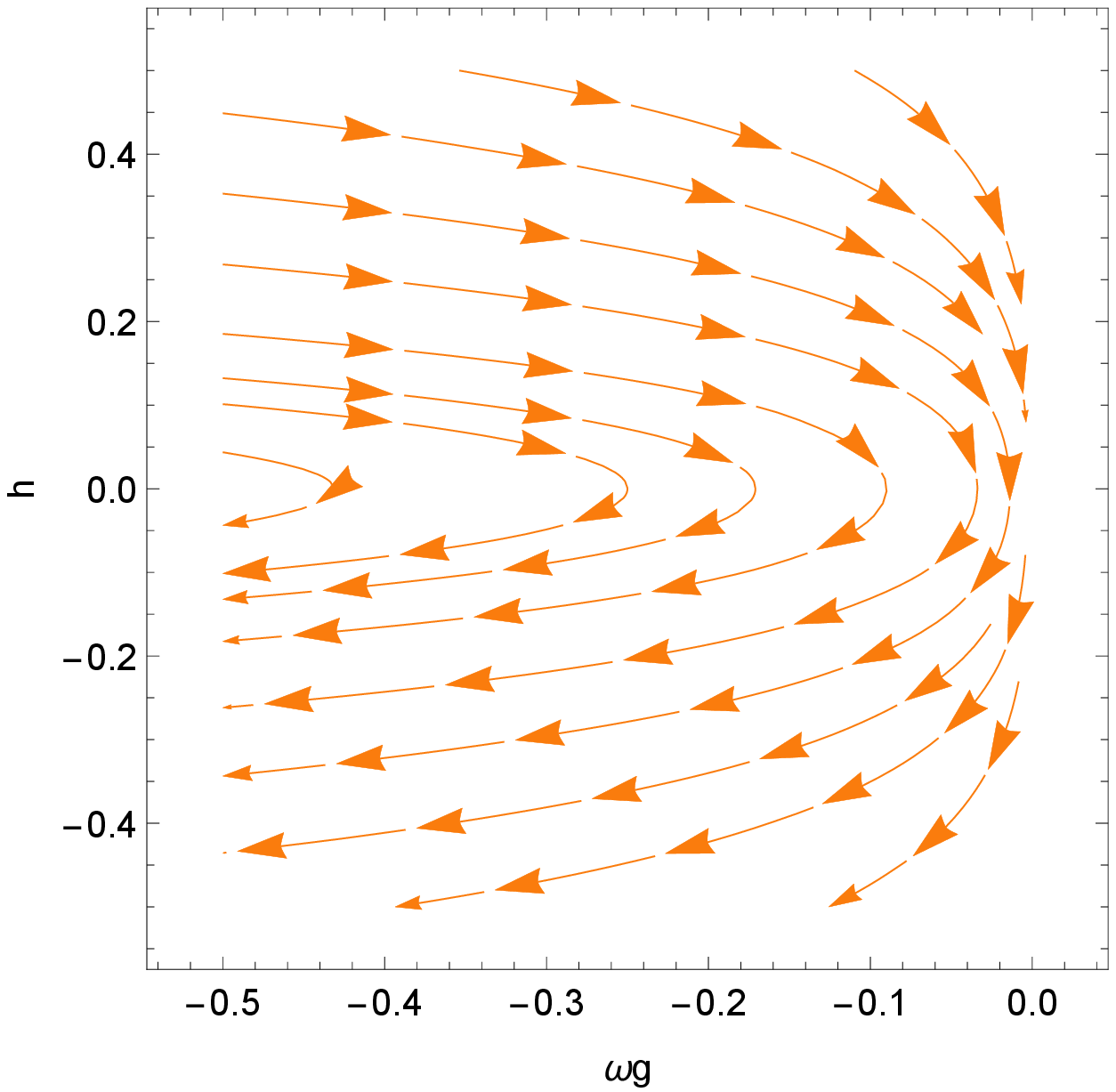}
%\mbox{\epsfxsize=14.2cm \epsffile{numericmodel2q.eps}}
\includegraphics[width=5.5cm]{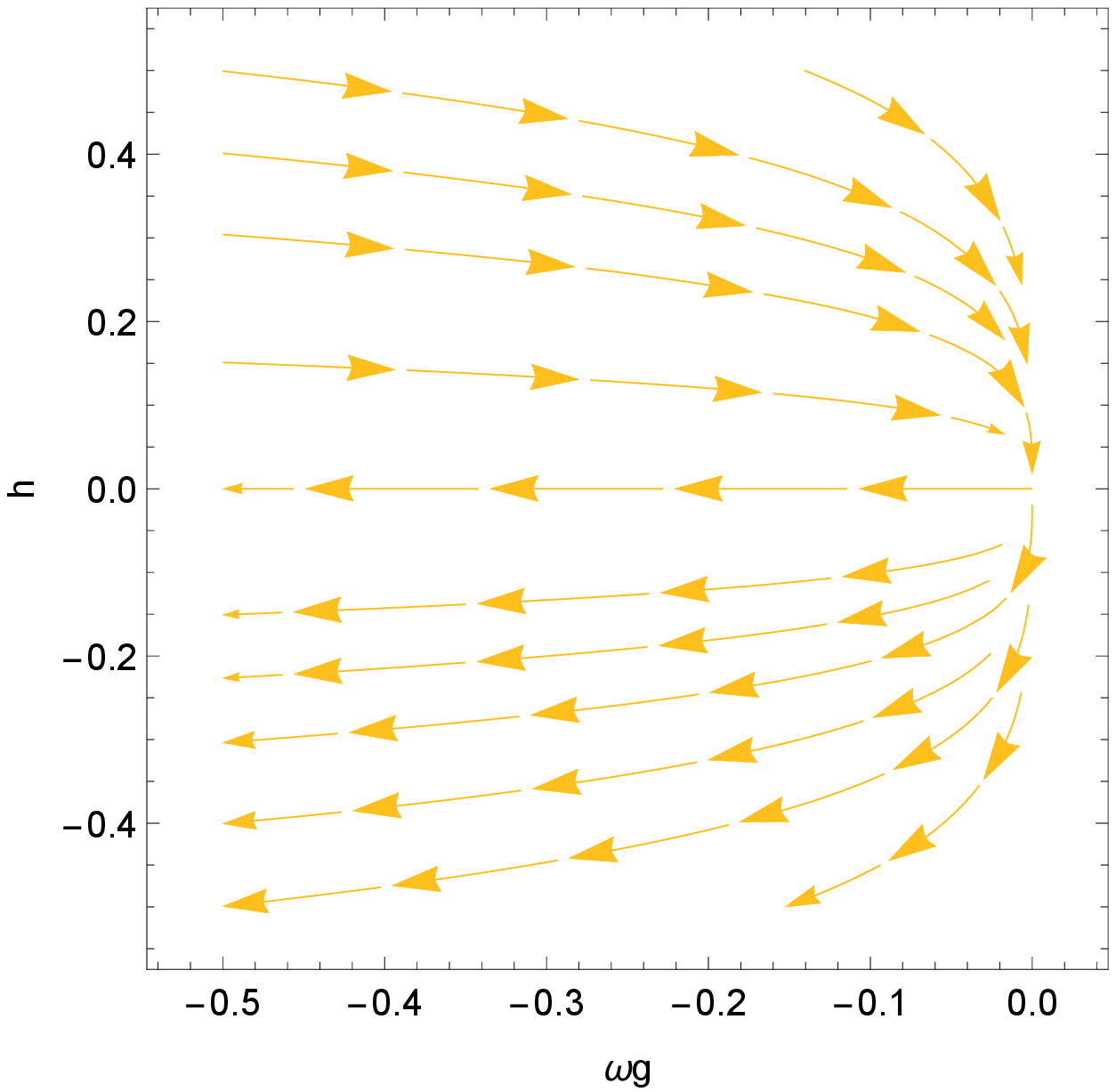}
%\mbox{\epsfxsize=14.2cm \epsffile{numericmodel2q.eps}}
\caption{Phase-space diagram for the dynamical system (\ref{se.01}%
)-(\ref{se.05}) in the $\omega_{g}-h$ surface~for $\gamma=\frac{4}{3}$%
,~$\beta=0~$and for three different values of $\omega_{m}$. }%
\label{pp.06}%
\end{figure}

\begin{figure}[ptb]
\includegraphics[width=5.5cm]{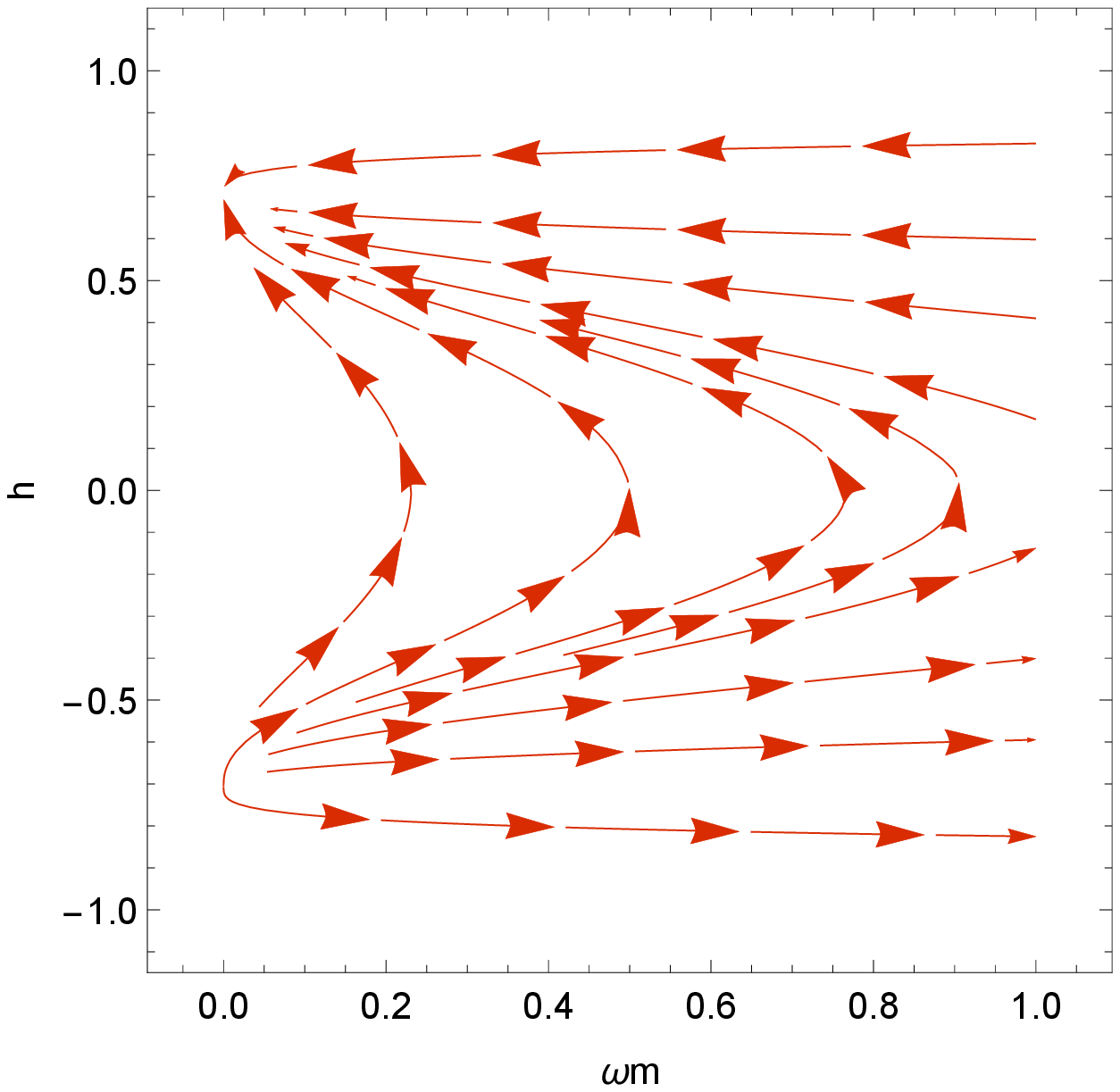}
%\mbox{\epsfxsize=14.2cm \epsffile{numericmodel2q.eps}}
\includegraphics[width=5.5cm]{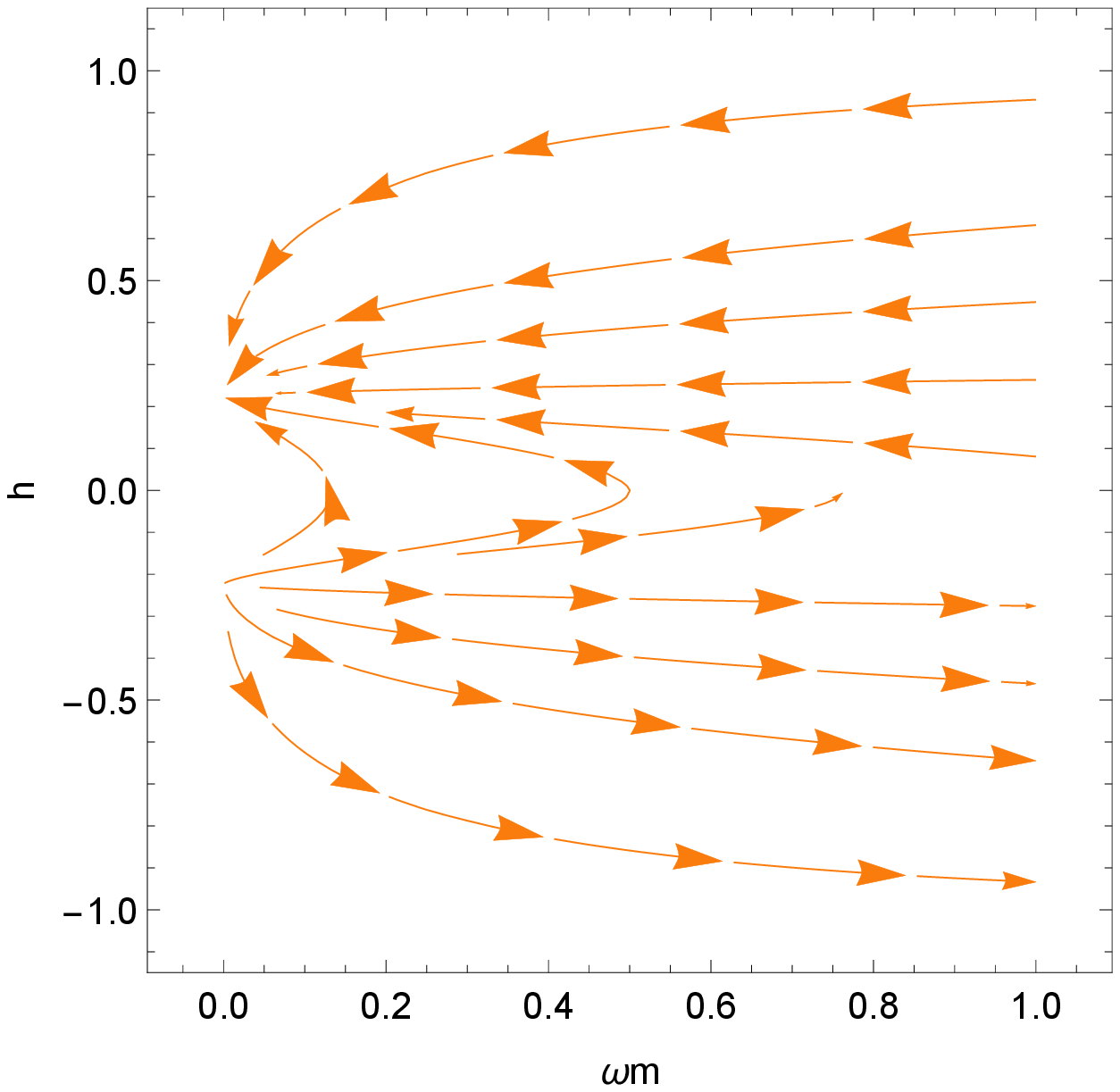}
%\mbox{\epsfxsize=14.2cm \epsffile{numericmodel2q.eps}}
\includegraphics[width=5.5cm]{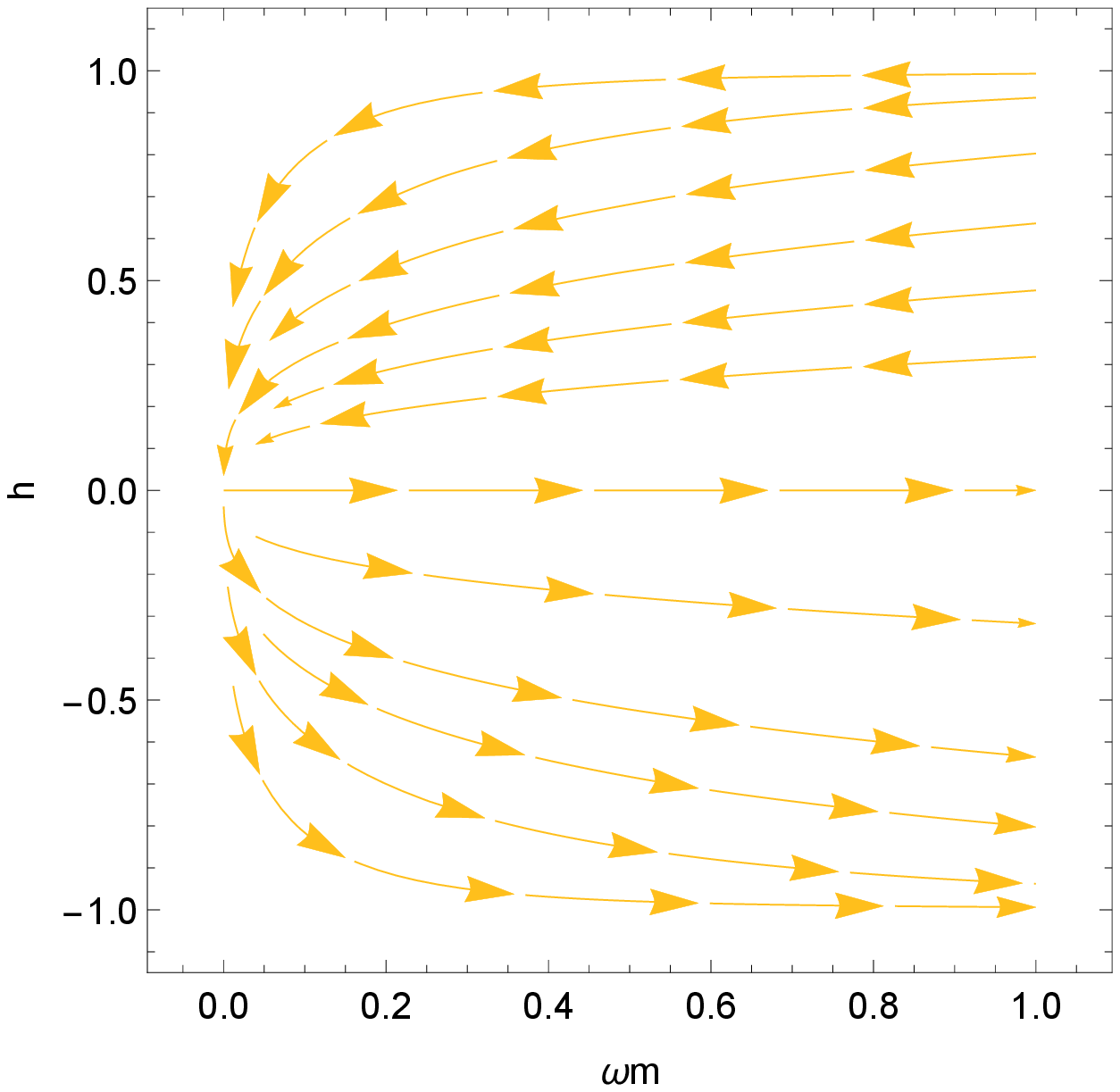}
%\mbox{\epsfxsize=14.2cm \epsffile{numericmodel2q.eps}}
\caption{Phase-space diagram for the dynamical system (\ref{se.01}%
)-(\ref{se.05}) in the $\omega_{m}-h$ surface,~for $\gamma=\frac{4}{3}%
$,~$\beta=0~$and for three different values of $\omega_{g}$. }%
\label{pp.05}%
\end{figure}

\begin{figure}[ptb]
\includegraphics[width=5.5cm]{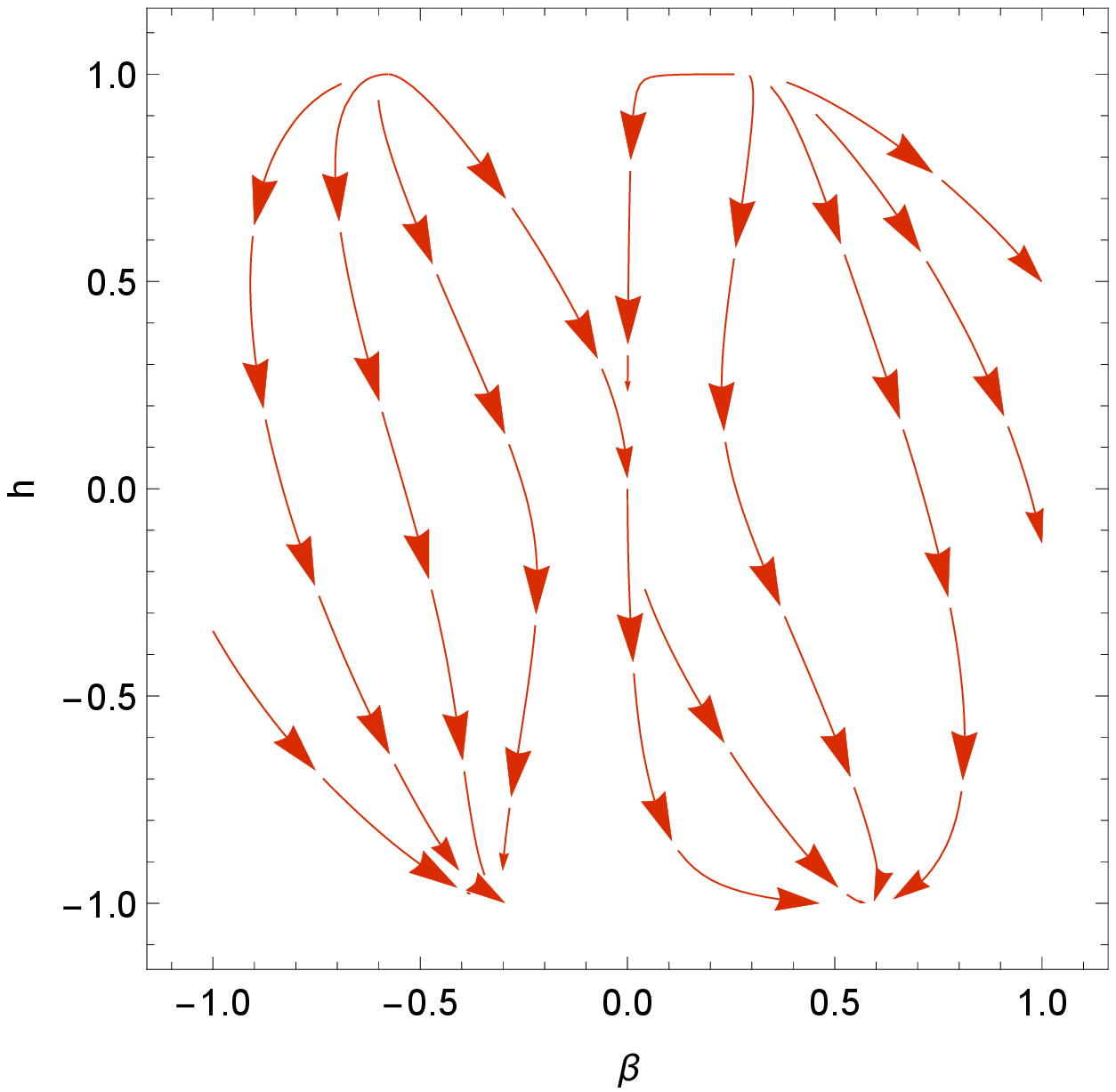}
%\mbox{\epsfxsize=14.2cm \epsffile{numericmodel2q.eps}}
\includegraphics[width=5.5cm]{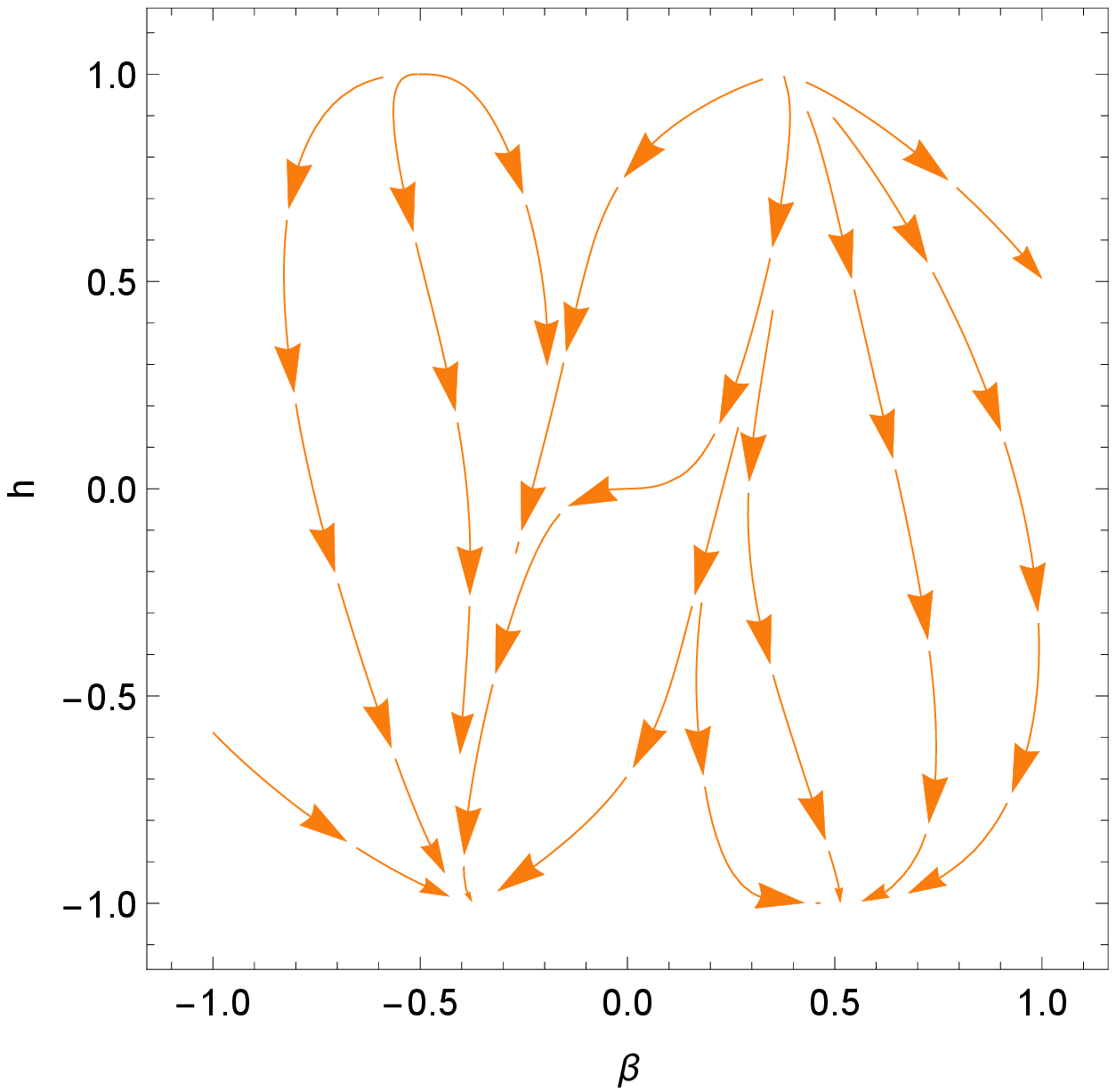}
%\mbox{\epsfxsize=14.2cm \epsffile{numericmodel2q.eps}}
\includegraphics[width=5.5cm]{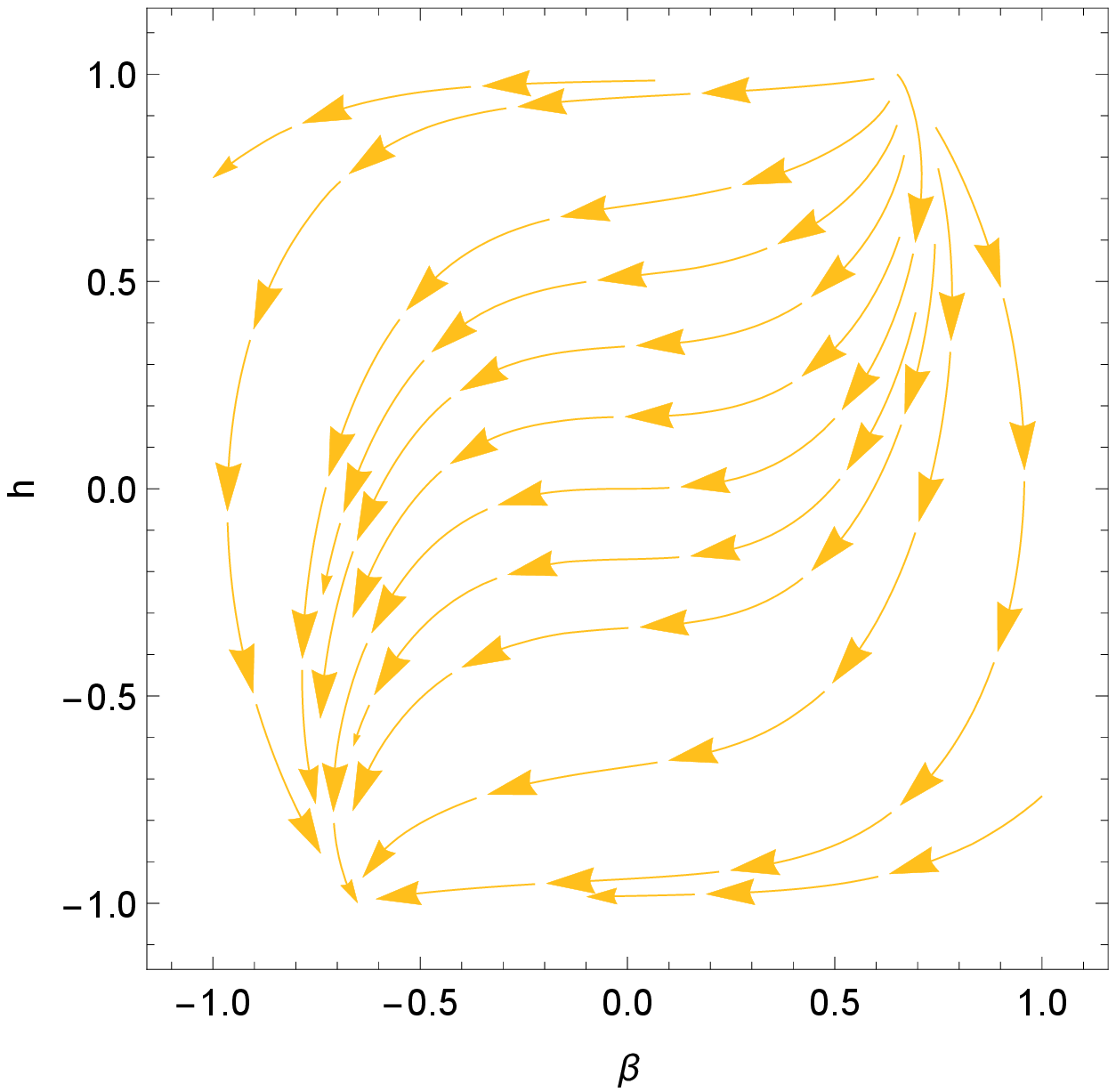}
%\mbox{\epsfxsize=14.2cm \epsffile{numericmodel2q.eps}}
\caption{Phase-space diagram for the dynamical system (\ref{se.01}%
)-(\ref{se.05}) in the $~\beta-h$ surface~for $\gamma=\frac{4}{3}$%
,~$\omega_{m}=-2\omega_{r},~\omega_{r}<-\frac{1}{2}~$and for three different
values of $\varepsilon$.}%
\label{pp.07}%
\end{figure}

Now we discuss the stability and the physical parameters of the points $P_{1}$
and $P_{2}$.

a. At point $P_{1}~$the anisotropic parameter $\beta~$ is not a constraint,
which means that $P_{1}$ describes a surface of critical points on the
phase-space. Since $\beta$ is unconstrained, point $P_{1}$ can describe
solutions in the Kantowski-Sachs family and in the FLRW-like family. From the
algebraic equation (\ref{se.06}), we can derive the parameter $\omega_{R}$,
namely,
\begin{equation}
\omega_{R}=\frac{9\beta^{2}}{2-\gamma}\left(  \gamma-\frac{4}{3}\right)  .
\end{equation}
Hence, the final geometry of the solution at $P_{1}~$depends upon the equation
of state parameter,$\gamma$, for the ghost field, $\rho_{g}$. If we assume
that $\beta\neq0$, then for $\gamma=\frac{4}{3}$ the solution at point $P_{1}$
describes a Bianchi I spacetime, for $\gamma>\frac{4}{3}$, the geometry is
that of Bianchi III, while, when $\gamma<\frac{4}{3},$~it follows that
$\omega_{R}<0$, which means that the $^{\left(  3\right)  }R<0$ and the
solution at point $P_{1}$ describes a Kantowski-Sachs universe. Furthermore,
at the special limit where $\beta=0$, $P_{1}$ describes the Minkowski
spacetime. We study the stability of the solution at $P_{1}$ in the
four-dimensional subspace $\left\{  \beta,\varepsilon,\omega_{m},\omega
_{g}\right\}  $ when $h\rightarrow0$. We find that there exists an eigenvalue,
positive real-valued, for the matrix which defines the linearized system.
Therefore, the solution at $P_{1}$ is unstable in the 4-dimensional subspace
$\left\{  \beta,\varepsilon,\omega_{m},\omega_{g}\right\}  $ and consequently
also in the 5-dimensional space in which the dynamical system evolves.
Two-dimensional phase-space diagrams are presented in Figures \ref{pp01}%
-\ref{pp04}, from which it is clear that $P_{1}$ describes an
unstable\ Einstein static solution. Moreover, from the phase-space diagrams we
observe that unstable oscillatory behaviours exist around $P_{1}.$ The figures
are for $\gamma=\frac{4}{3}$, Figs. \ref{pp01} and \ref{pp02} are in the
surface $\omega_{m}-h$,~$\omega_{g}-h$, respectively, Fig. \ref{pp03} is in
the surface $\beta-\varepsilon$ and vectors in Fig. \ref{pp04} are on the
surface $\beta-h$. 

b. The solution at point $P_{2}$ describes an isotropic static universe
because $\beta=\varepsilon=0$, and more specifically it is the inhomogeneous
FLRW space with positive spatial curvature, i.e., $\omega_{R}=-\omega_{g}$. We
remark that $P_{2}$, like point $P_{1}$, is actually a surface -- a family of
solutions where~$\omega_{m}=-2\omega_{g}$ but with $\omega_{R}=\frac
{\omega_{m}}{2}$, which means that the spatial 3-curvature is positive. In
order to study the stability of the solution we calculate the eigenvalues of
the linearized system and they are%
\[
e_{1}=0~,~e_{2}^{\pm}=\pm\sqrt{\frac{\gamma\omega_{g}}{3}}~,~e_{3}^{\pm}%
=\pm\sqrt{\frac{\left(  2-\gamma\right)  \left(  -\omega_{g}\right)  }{6}}\,.
\]
Hence, there exists always a positive eigenvalue and so we can infer that the
solution at $P_{2}$ is unstable. However, one of the eigenvalues has nonzero
imaginary part (because $\omega_{m}=-2\omega_{g}$) which means that periodic
behaviour exists. In particular the imaginary eigenvalues are in the
$\omega_{g}-h$ surface, and indeed periodic behaviour is observed in Fig.
\ref{pp.06}. This means that small perturbations around $P_{2}$ in the
$\omega_{g}-h$ surface give a behaviour similar to that of the solution
(\ref{ee.22aa}). In Figures \ref{pp.05} and \ref{pp.07} the phase-space
diagrams in the $\omega_{m}-h$ and $\beta-h$ surfaces are presented,
respectively. 

\section{Conclusions}

\label{section5}

We have considered the Szekeres dust universe with an additional homogeneous
and isotropic ghost field. The equation of state parameter for the ghost field
was assumed to be $p_{g}=\left(  \gamma-1\right)  \rho_{g}$ and $\rho_{g}<0$.
We were able to simplify the gravitational field equations and determine the
existence of two possible families of solutions. Unlike in the absence of the
ghost field, the first family of solutions describes spatially homogeneous
Kantowski-Sachs universes, while the second family of solutions describes
inhomogeneous FLRW-like universes. The specific forms of the spacetimes are
similar to those determined in the case of an homogeneous scalar field and
dust in the Szekeres metrics\cite{jdbsc}. However, the existence of the ghost
field produces new possible behaviours for the scale factors of these
universes. Specifically, it is possible to have Einstein-static solutions in
the Kantowski-Sachs family while a cyclic solution was found analytically for
the FLRW-like family of spacetimes.

By studying the critical points of the gravitational field equations expressed
in terms of the kinematic quantities we have found two points which describe
Einstein static solutions, points $P_{1}$ and $P_{2}$, which are sources. More
specifically, $P_{1}$ and $P_{2}$ actually describe surfaces in the dynamical
phase-space: $P_{1}$ exists for both of the families while $P_{2}$ describes
an Einstein static solution in the FLRW family of solutions. While the
Einstein solutions are unstable, from the numerical simulations it is easy to
observe that for specific initial conditions around the critical points cyclic
behaviour appears which is agreement with the cyclic solution determined
analytically. These are the first studies, via exact solutions, of
inhomogeneous oscillating universes. We have not introduced dissipative
processes but entropy production could be introduced in order to study the
evolution of cycle size and length as the universe evolves through successive
maxima \cite{tol, BD,jkm,jkm2}

\begin{acknowledgments}
JDB is supported by the Science and Technology Facilities Council (STFC) of
the United Kingdom. AP acknowledges financial supported of FONDECYT grant No. 3160121.
\end{acknowledgments}

\end{document}